\documentclass[pra,twocolumn,superscriptaddress,floatfix,showpacs]{revtex4}
\usepackage{graphicx}
\usepackage{bm}
\usepackage{amsmath}
\usepackage{amssymb}

\newcommand{\eqname}[1]{\label{eq:#1}}
\newcommand{\bgar}{\begin{eqnarray}}
\newcommand{\enar}[1]{\label{eq:#1}\end{eqnarray}}

\newcommand{\kk}{ {\bf k}}
\newcommand{\rr}{ {\bf r}}

\newcommand{\qq}{ {\bf q}}

\newcommand{\Real}[1]{\textrm{Re}\left[#1\right]}
\newcommand{\Imag}[1]{\textrm{Im}\left[#1\right]}
\newcommand{\eq}[1]{(\ref{eq:#1})}

\newcommand{\Psihd}{\hat\Psi^\dagger}

\newcommand{\Psih}{\hat\Psi}

\newcommand{\ahd}{\hat a^\dagger}
\newcommand{\ah}{\hat a}

\newcommand{\chd}{\hat c^\dagger}

\newcommand{\ch}{\hat c}

\begin{document}

\title{A semi-classical field 
method for the equilibrium Bose gas and application to
thermal vortices in two dimensions}

\author{Luca Giorgetti}
\affiliation{CNR-BEC-INFM and Dipartimento di Fisica, Universit\`a
di Trento, I-38050 Povo, Italy}

\author{Iacopo Carusotto}
\affiliation{CNR-BEC-INFM and Dipartimento di Fisica, Universit\`a
di Trento, I-38050 Povo, Italy}

\author{Yvan Castin}
\affiliation{Laboratoire Kastler Brossel, Ecole normale sup\'erieure, 
UPMC, CNRS, 24 rue Lhomond,
F-75231 Paris Cedex 5, France}

\begin{abstract}
We develop a semi-classical field method for the study of the weakly
 interacting Bose gas at finite temperature, which, 
contrarily to the usual classical field model,
does not suffer from an ultraviolet cut-off dependence.
We apply the method to the study of thermal vortices in spatially
 homogeneous, two-dimensional systems.
We present numerical results for the vortex density and the vortex pair  
 distribution function.
Insight in the physics of the system is obtained by comparing the
 numerical results with the predictions of simple analytical models.
In particular, we calculate
the activation energy required to form a vortex pair
at low temperature.
\end{abstract}

\date{\today} \maketitle

\section{Introduction}

Classical field theories are a widespread and flexible tool for the
study of many aspects of the physics of ultracold Bose gases.
Developed in particular to address time-dependent problems related to dynamical
aspects of the Bose-Einstein phase 
transition~\cite{CFT-dyn,Sachdev}
they can also be used to study thermal equilibrium properties 
of the weakly interacting Bose gas \cite{Burnett,Rzazewski0}.
A major example is the quantitative prediction
of the shift of the Bose-Einstein condensation temperature 
due to atomic interactions~\cite{Baym} that has been obtained by
means of a Monte Carlo sampling of a classical field model~\cite{3DDeltaT}:
as long as the physics of the system is determined by the low-energy
modes, classical field models provide reliable results on the full
quantum problem. 
Classical field techniques have also been applied to obtain
analytical and numerical predictions for reduced dimensionality 
Bose systems~\cite{Scalapino,CFT,Rzazewski0,Hutch}, 
including the calculation of the critical temperature
for the Berezinskii-Kosterlitz-Thouless transition in two
dimensions~\cite{Svistunov_tc2da,Svistunov_tc2db}.
However, an ultraviolet cut-off has to be introduced in most of these
classical field techniques in order to avoid ultraviolet divergences
analogous to the blackbody catastrophe of classical statistical
mechanics, and this raises the problem of a possible cut-off
dependence of some of the physical results.  

On the other side, several exact reformulations of the many boson problem
have been developed.
Although they have successfully served as a starting point for Quantum
Monte Carlo simulations~\cite{QMC,worm} of the thermal properties of
Bose systems such as liquid Helium and ultracold atomic gases 
\cite{Krauth,Ceperley2,stat_N0}, they
often lack the intuitiveness of classical field theories where the
physics is described in terms of a simple distribution function in the
functional space of c-number fields. 

The present paper is devoted to the development, the validation, and the
first application of a {\em semi}-classical field theory which tries to
combine a regular behavior in the ultraviolet limit with a transparent 
intuition of the physics of the system.
As in classical field theories, the density matrix of the Bose system is
written in terms of a distribution in the space of c-number
fields.
In the semi-classical theory, this distribution is however much more
complex than a simple Boltzmann factor $\exp(-E/k_B T)$, where $E$ would be 
the Gross-Pitaevskii energy of the field configuration, and has to be
obtained as the result of an imaginary-time Gross-Pitaevskii evolution
starting from an initially uniform distribution in functional space.

A first application of the method is then presented to the study of 
thermal vortices in a homogeneous two-dimensional Bose gas, in
particular their density and their pair distribution
function. Experimentally, the two-dimensional Bose gas has been realized
some time ago~\cite{Safonov,2D_atoms}, but it is only recently that
several experiments have given indications of the presence
of vortices in finite temperature samples~\cite{Dalib_vort,Dalibard2D,Cornell_APS}, 
and this raises the question of the link between observable quantities
(e.g the vortex density), and theoretical concepts such as the
Berezinskii-Kosterlitz-Thouless (BKT)
transition~\cite{BKT,Minnhagen,Svistunov_tc2da,Svistunov_tc2db,Markus}. 
Most of the existing theoretical treatments neglect all density
fluctuations other than the ones in the vicinity of a vortex core, and
eventually map the 2D Bose gas problem 
onto the XY model of statistical mechanics~\cite{XY}.
Although this approximation is expected to provide a good description
of atomic gases trapped in 2D optical
lattices~\cite{Cornell_APS,Trombetta_2D,Trombetta_2D_BEC}, it seems far 
from being accurate for 
spatially continuous systems: at temperatures of the order of
the BKT transition temperature, the amplitude of the density
fluctuations in the gas is not negligible as compared to the density
itself \cite{Svistunov2D}.   
Our work aims at going beyond this approximation so to fully
include the effect of density fluctuations.
The fact that it is based on c-number fields gives to the
present semi-classical method an advantage over 
standard Quantum Monte Carlo techniques in view of the study of
vortices.

The paper is divided in two main parts.
In the first part (Sec.\ref{sec:SC}), we introduce the semi-classical
method in the grand-canonical (Sec.\ref{sec:GC}) and in the canonical
(Sec.\ref{subsec:itce}) ensembles, and we characterize its range of
applicability (Sec.\ref{sec:limits}). 
In the second part (Sec.\ref{sec:appl}), we discuss the physics of the
two-dimensional Bose gas.
The numerical results are presented in
Sec.\ref{sec:numerical}: different observables are considered,
e.g.\ the normal and non-condensed fractions, the density 
fluctuations, the vortex density, and the vortex pair-distribution
function.
In Sec.\ref{subsec:Bog} the effect of Bose condensation on the vortex density
in the finite size ideal gas is discussed analytically; 
this requires the use of the canonical
ensemble, which introduces new features with respect to the well-studied grand canonical
case \cite{Halperin,Berry}.
In Sec.\ref{sec:analyt} a simple model including the interacting case
is developed to understand the numerical
results, principally the ones for the vortex density $n_{v,+}$: 
an activation law
of the form $n_{v,+}\propto \exp(-\Delta/k_B T)$ is found in the
low-temperature regime, and the dependence of $\Delta(T)$ on the
system parameters such as the interaction strength and the system size is
discussed: the main qualitative differences between the ideal and the
interacting gas behaviors are pointed out.
Conclusions are finally drawn in Sec.\ref{sec:Conclu}.

\section{The semi-classical method}
\label{sec:SC}

\subsection{In the grand-canonical ensemble}
\label{sec:GC}

Consider a Bose field defined on an square lattice of ${\mathcal N}$
points with periodic boundary conditions; $V$ is the total volume of the
quantization box and $dV=V/{\mathcal N}$ is the volume of the unit cell of the
lattice.
The Bose field operators $\Psih(\rr)$ obey the Bose commutation
relations $[\Psih(\rr),\Psihd(\rr')]=\delta_{\rr,\rr'}/dV$.

The state of the Bose field is described by the density operator $\rho$,
which can be expanded in the so-called Glauber-P representation on coherent
states: 
\begin{equation}
\rho=\int\!{\mathcal D}\psi\,P[\psi]\,|\textrm{coh}:\psi\rangle
\langle\textrm{coh}:\psi|,
\eqname{exp_coh}
\end{equation}
where $P[\psi]$, the Glauber-P distribution, is guaranteed to exist 
in the sense of distributions but in general is not a positive nor even 
a regular function \cite{Glauber,quantum_optics,quantum_noise}.
$\psi(\rr)$ is here a c-number field defined on the
lattice, the coherent state is defined as usual as:
\begin{equation}
|\textrm{coh}:\psi\rangle=\exp\left[-\frac{1}{2}\,\|\psi\|^2\right]\,
\exp\left\{\sum_\rr\!dV\,\psi(\rr)\,\Psihd(\rr)\right\}\,|0\rangle,
\end{equation}
where $||\psi||^2=dV \sum_{\mathbf{r}} |\psi(\mathbf{r})|^2$,
and the functional integration is performed over the value of the
complex field at each of the ${\mathcal N}$ sites of the lattice: 
\begin{equation}
{\mathcal D}\psi=\prod_\rr d\Real{\psi(\rr)}\,d\Imag{\psi(\rr)}.
\end{equation}

The homogeneous Bose gas is described by the following
second-quantized Hamiltonian:
\begin{multline}
{\mathcal H}=
%\sum_\rr\Psihd(\rr)\,[h_0-\mu]\,\Psih(\rr)+
\sum_\kk \left[\frac{\hbar^2 k^2}{2m}-\mu\right] \ahd_\kk \ah_\kk \\
+\frac{g_0}{2} \sum_\rr dV\,\Psihd(\rr)\Psihd(\rr)\Psih(\rr)\Psih(\rr).
%+\sum_\rr dV\,U_{\rm ext}(\rr) \Psihd(\rr)\Psih(\rr),
\eqname{Hamilt}
\end{multline}
The single-particle dispersion relation within the first Brillouin zone
is taken as parabolic with mass $m$, 
$\mu$ is the chemical potential, and the interactions are modeled by a
two-body discrete delta potential of strength $g_0$.

The gas is assumed to be at thermal equilibrium at a
temperature $T$, so that the unnormalized density operator is 
$\rho_{\rm eq}(\beta)=\exp[-\beta\,{\mathcal H}]$ 
with $\beta=1/k_B T$.
This density operator can be obtained as the result of an imaginary-time
evolution: 
\begin{equation}
\frac{d\rho_{\rm eq}}{d\tau}=-\frac{1}{2}\{ {\mathcal H},\rho_{\rm eq}
\}=-\frac{1}{2} [ {\mathcal H}\rho_{\rm eq}+\rho_{\rm eq}{\mathcal H}]
\end{equation}
during the ``time'' interval $\tau=0\rightarrow \beta$, starting from the 
identity operator $\rho_{\rm eq}(\tau=0)={\mathbf 1}$.

In the Glauber-P representation, the imaginary-time
evolution takes the form of a Fokker-Planck-like partial
differential equation: 
\begin{eqnarray}
\partial_\tau P[\psi]&=&-E[\psi]\,P[\psi]
-\sum_\rr
\Big[\partial_{\psi(\rr)}\left(F[\psi]\,P[\psi]\right)
 \nonumber \\
&&+
\frac{g_0}{4 dV}\partial^2_{\psi(\rr)}(\psi^2(\rr)\,P[\psi])+
\textrm{c.c.}\Big]
\eqname{FP}
\end{eqnarray}
for the distribution function $P[\psi]$ in the phase-space of the 
c-number fields defined on the lattice.
The derivatives with respect to the complex field $\psi(\rr)$ are
defined as usual as:
\begin{equation}
\partial_{\psi(\rr)}=\frac{1}{2}\left[\partial_{\Real{\psi(\rr)}}
-i \partial_{\Imag{\psi(\rr)}} \right].
\end{equation}

The first term in the right-hand side of \eq{FP} acts on the weight of the
wavefunction $\psi$ and involves the mean-field energy of the complex
field $\psi(\rr)$: 
\begin{equation}
E[\psi]=
\sum_\rr dV\,\psi^*(\rr)\,[h_0-\mu]\,\psi(\rr)+\frac{g_0}{2}\sum_\rr
dV\,|\psi(\rr)|^4 .
\eqname{weight}
\end{equation}
$h_0$ is a shorthand for the single-particle Hamiltonian, whose
$k$-space form is $h_0=\hbar^2 k^2/(2m)$.

The second term is a drift term consisting of the imaginary-time
Gross-Pitaevskii evolution:  
\begin{equation}
F[\psi](\rr)=-\frac{1}{2 dV}\,\partial_{\psi^*(\rr)}E[\psi]=
-\frac{1}{2}\left[h_0-\mu+g_0\,|\psi(\rr)|^2\right]\,\psi(\rr).
\eqname{drift}
\end{equation}
Finally, the diffusion terms involving the second-order
derivatives are local in space, but have a non-positive-definite
diffusion matrix:
\begin{equation}
D(\rr)=-\frac{g_0}{4 dV}
\left(
\begin{array}{cc}
0 & \psi^2(\rr) \\
\psi^{*2}(\rr) & 0
\end{array}
\right).
\eqname{diff}
\end{equation}
A complete solution of the partial differential equation \eq{FP} would provide
the exact result of the lattice quantum field problem defined by the Hamiltonian
\eq{Hamilt}.
Unfortunately, the non-positive-definite nature of the diffusion matrix
\eq{diff} prevents the Fokker-Planck-like equation \eq{FP} 
from being directly mappable on a
stochastic field problem for $\psi$. 
Some approximation schemes are therefore required in order to perform
numerical simulations within the Glauber-P framework. 

In our previous work~\cite{CFT}, the high-temperature physics of the
one-dimensional Bose gas 
was studied by keeping only the first term in the right-hand side of \eq{FP}. 
The resulting distribution in the phase-space of the
 c-number fields is the usual Boltzmann one 
$P[\psi]=\exp(-E[\psi]/k_B T)$ in terms of the mean-field energy
\eq{weight}. 
A better approximation is obtained by keeping also the drift force
\eq{drift} and neglecting the diffusion term \eq{diff} only.
In this case, the partial differential equation \eq{FP} can be mapped
onto a deterministic evolution for the field $\psi(\rr)$ and a
weight ${\mathcal W}$:   
\begin{eqnarray}
\partial_\tau \psi(\rr,\tau)&=& -\frac{1}{2}[h_0-\mu+g_0\,|\psi(\rr,\tau)|^2]\,\psi(\rr,\tau)
 \eqname{phi}, \\
\partial_\tau {\mathcal W}(\tau)&=&-E[\psi(\tau)]\,{\mathcal W}(\tau). \eqname{W}
\end{eqnarray}
Physical quantities are then obtained as averages over the initial
values for $\psi$. A possible representation of the initial state
$\rho_{\rm eq}(\tau=0)={\mathbf 1}$ is to take the initial value of the field
$\psi(\rr,\tau=0)$ at each lattice point as uniformly distributed in the
complex space and to take a constant initial weight $\mathcal{W}(\tau=0)=w_0$.
This leads to the {\em semi-classical} approximation for the density operator
at temperature $T$:
\begin{equation}
\rho_{\rm SC} = \int \mathcal{D}\psi(0) \, \mathcal{W}(\beta)
|\mathrm{coh}:\psi(\beta)\rangle \langle \mathrm{coh}:\psi(\beta)|,
\eqname{rho_SC_gcan}
\end{equation}
where both $\mathcal{W}(\beta)$ and $\psi(\beta)$ depend on the initial
value of the field $\psi(0)$.

As the diffusion term \eq{diff} is proportional to the interaction
strength $g_0$, the semi-classical approximation
becomes exact in the case of the free Bose field, i.e. for an ideal Bose
gas. As a consequence, it does not suffer 
from the typical ultraviolet divergences of classical field theories, 
even in presence of interactions.

\subsection{Limits of validity}
\label{sec:limits}
In order to validate the semi-classical approximation and appreciate its power
and its limits, it is interesting to apply it to the simple case of
the Bogoliubov Hamiltonian
\begin{equation}
{\mathcal H}_{\rm Bog}=\sum_{\kk\neq \mathbf{0}} \left(\frac{\hbar^2 k^2}{2m}+\mu\right)\, \ahd_\kk
\ah_\kk  
+\frac{\mu}{2} \left(\ahd_\kk \ahd_{-\kk}+\ah_\kk \ah_{-\kk}\right).
\eqname{H_Bog}
\end{equation}
This Hamiltonian being quadratic in the field operators, the
semi-classical equations (\ref{eq:phi}-\ref{eq:W}) can be
analytically solved and their prediction compared to the exact quantum
results.

By defining the operators $\ch_{\kk,+}=(\ah_\kk + \ah_{-\kk})/\sqrt{2}$
and $\ch_{\kk,-}=(\ah_\kk - \ah_{-\kk})/(i\sqrt{2})$, the Bogoliubov
Hamiltonian \eq{H_Bog} can be rewritten as a sum of terms involving
independent $\kk$ modes:
\begin{multline}
{\mathcal H}_{\rm Bog}={\sum_{\kk,\epsilon=\pm}\!\!}'\, {\mathcal H}_{\kk,\epsilon}={\sum_{\kk,\epsilon=\pm}\!\!}'\,
  \left(\frac{\hbar^2 
  k^2}{2m} +\mu\right) \chd_{\kk,\epsilon} \ch_{\kk,\epsilon} \\
+\frac{\mu}{2} \left(\chd_{\kk,\epsilon}
  \chd_{\kk,\epsilon}+ \ch_{\kk,\epsilon} \ch_{\kk,\epsilon}\right).
\eqname{Bogo}
\end{multline}
In this way, the Glauber-P distribution factorises as a product of
independent factors involving the different $\kk$ modes.
To avoid double-counting of modes, the primed sum is restricted to those
$\kk$ vectors which are contained in an (arbitrarily chosen)
half-space. 

Each term of the Hamiltonian \eq{Bogo} has the simple structure of a
one-mode squeezing Hamiltonian:
\begin{equation}
\mathcal{H}_1=(E_k+\mu)\, \chd \ch +
\frac{\mu}{2}\,\left(\ch^2+\ch^{\dagger 2}\right),
\eqname{Hquad}
\end{equation}
with the kinetic energy coefficient $E_k=\hbar^2 k^2/(2m)$ and the $\ch$
operator corresponding to any of $\ch_{\kk,\pm}$ in \eq{Bogo}. 

Since the Hamiltonian \eq{Hquad} is quadratic, the exact Glauber-P
distribution for the thermal equilibrium state can be analytically 
obtained by means
of standard techniques \cite{quantum_noise}, as well as its
semi-classical approximation:
as shown in the Appendix \ref{appen:omm},
both distributions have a Gaussian form,
\begin{equation}
P(\gamma)  \propto e^{-(\mathrm{Re}\,\gamma)^2/\sigma_R^2} e^{-(\mathrm{Im}\,\gamma)^2/\sigma_I^2}.
\eqname{gf}
\end{equation}

\begin{figure}[htbp]
\includegraphics[width=8cm,clip]{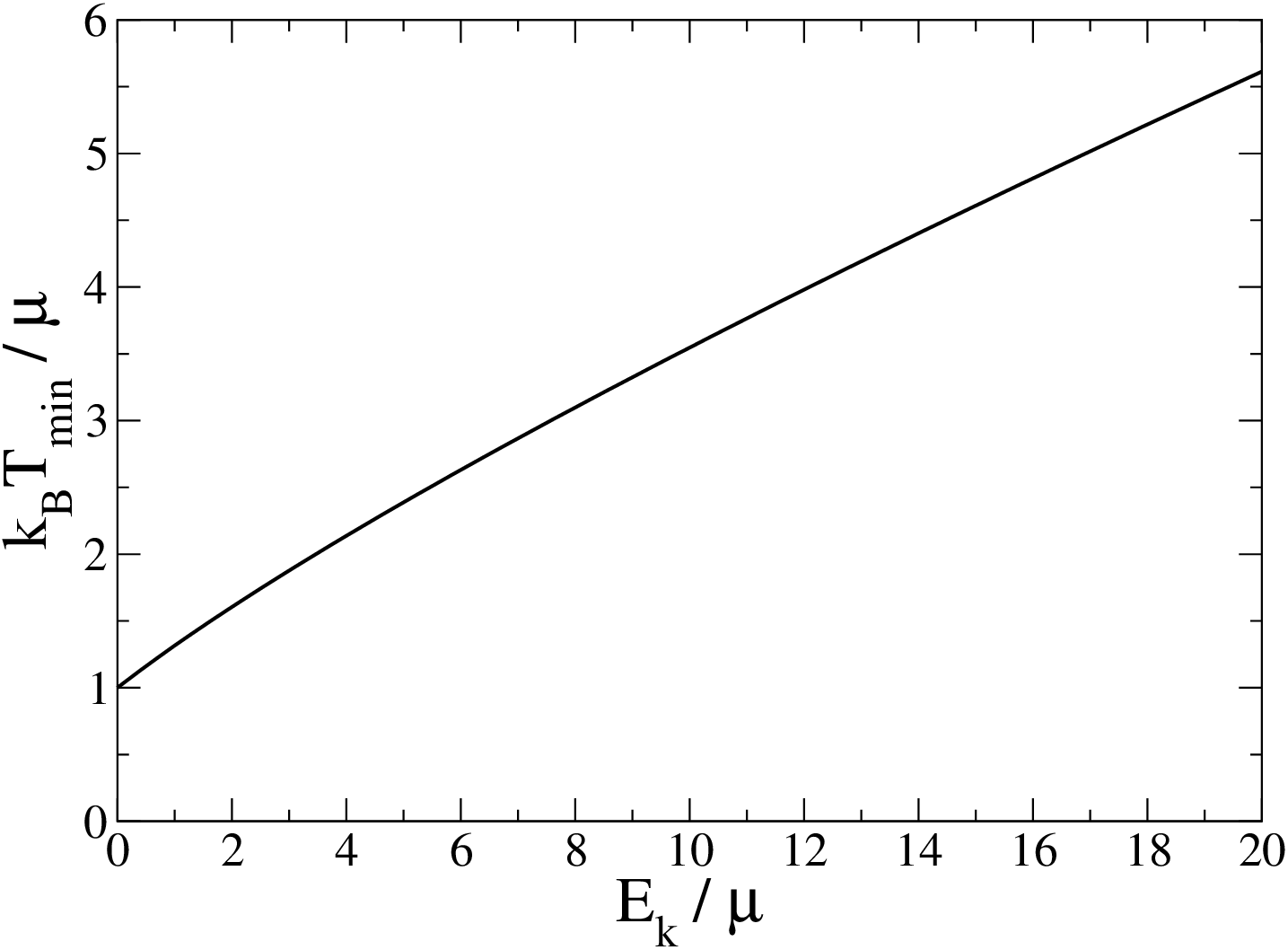}
\caption{In the Bogoliubov model, 
minimal value of the temperature $T_{\rm min}$
ensuring regularity and positivity of the Glauber-P distribution in
a mode $\mathbf{k}$, as a function of the kinetic energy coefficient $E_k$
of the mode.
\label{fig:T_min}}
\end{figure}

The widths for the exact distribution are given by
\begin{eqnarray}
\label{eq:srq}
\left(\sigma_R^2\right)_{\rm ex} &=&
\frac{1}{2}\left[ \left(\frac{E_k}{E_k+2\mu}\right)^{1/2} 
\!\!\mbox{cotanh}\left(\frac{\beta \epsilon_k}{2}\right)
-1\right] \\
\left(\sigma_I^2\right)_{\rm ex} &=&
\frac{1}{2}\left[ \left(\frac{E_k+2\mu}{E_k}\right)^{1/2} 
\!\!\mbox{cotanh}\left(\frac{\beta \epsilon_k}{2}\right)
-1\right] 
\end{eqnarray}
where $\epsilon_k=
[E_k(2\mu+E_k)]^{1/2}$ is the energy of the Bogoliubov mode.
When the temperature is too low, $(\sigma_R^2)_{\rm ex}$ becomes negative, so that the Glauber-P
distribution ceases to exist as a regular function \cite{quantum_optics,quantum_noise}.
The corresponding lower bound on the temperature is plotted in
Fig.\ref{fig:T_min}. Two limiting cases are easily isolated: for
low-energy modes such that $E_k\rightarrow 0$, the positivity condition 
for the Glauber-P distribution is the
simple one $k_B T > \mu$. For high energy modes, the condition is
instead more stringent, $k_B T > (E_k+\mu)/|\log(\mu/2E_k)|$.

The widths for the semi-classical approximation are given by
\begin{eqnarray}
\label{eq:srsc}
\left(\sigma_R^2\right)_{\rm SC} &=& \left[e^{\beta(E_k+2\mu)}-1\right]^{-1} \\
\left(\sigma_I^2\right)_{\rm SC} &=&\left[e^{\beta E_k}-1\right]^{-1}.
\label{eq:sisc}
\end{eqnarray}
As expected, they remain positive at all temperature.

\begin{figure}[htbp]
\includegraphics[width=8cm,clip]{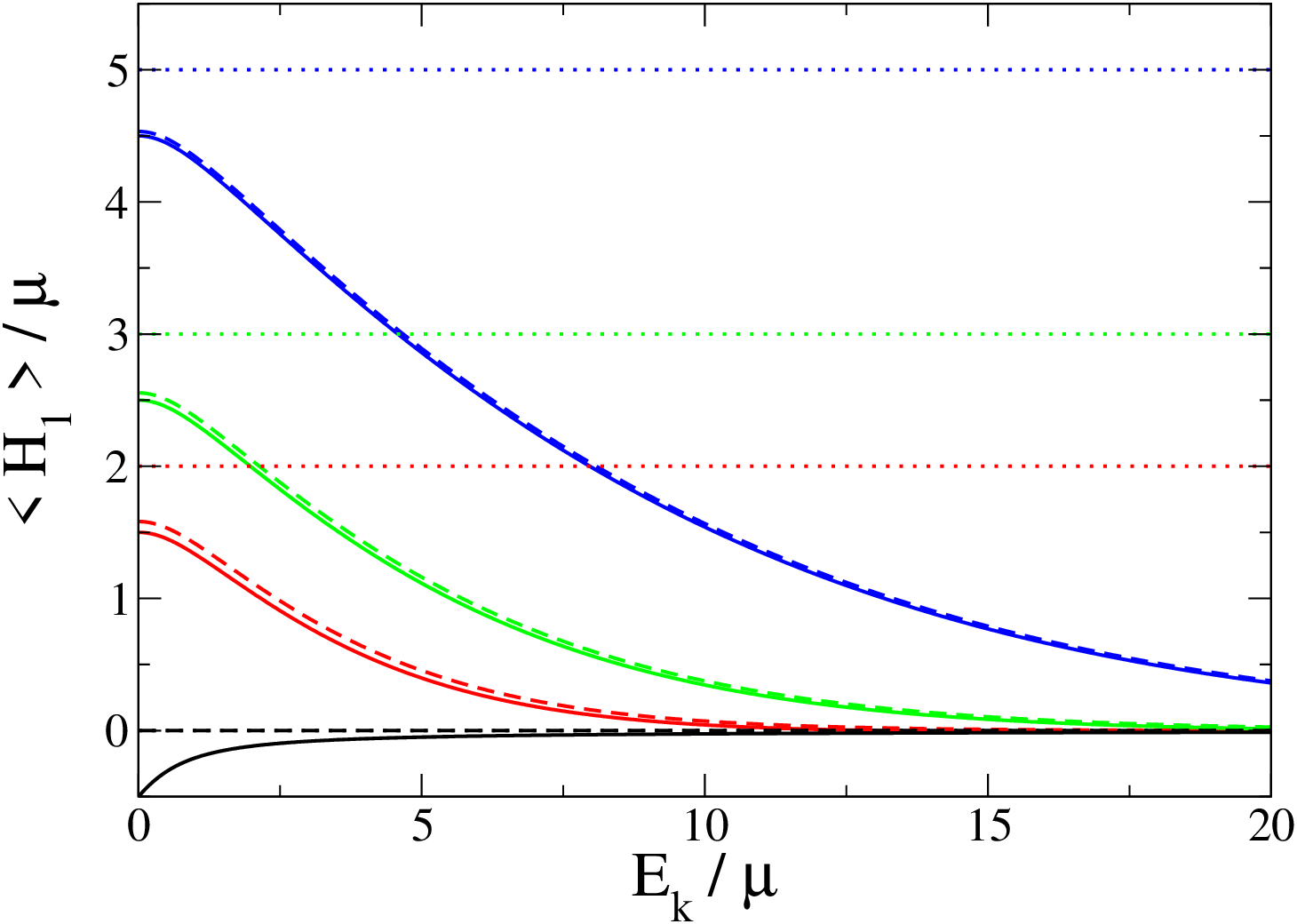}
\caption{(Color online) In the Bogoliubov model,
mean energy in a mode as a function of the mode kinetic energy coefficient
$E_k$ for different values of the temperature $k_B T/\mu =0, 2, 3, 5$ (from bottom to
top). Solid lines: quantum result. Dashed lines: semi-classical
theory. Dotted lines: classical field approximation.
\label{fig:Bogo_en}}
\end{figure}

These results are the starting point for detailed comparison of the
semi-classical predictions to the exact quantum results for the most
significant observables. Let us start with the {\em mean energy}.
The semi-classical value is:
\begin{equation}
\langle \mathcal{H}_1 \rangle_{SC}=
\frac{1}{2}\left[
\frac{E_k+2\mu} {e^{\beta(E_k+2\mu)}-1 } 
+
\frac{E_k} {e^{\beta E_k}-1 } 
\right],
\eqname{Esc}
\end{equation}
which is to be compared to the exact value
\begin{equation}
\langle  \mathcal{H}_1 \rangle_{\rm ex}=\frac{\epsilon_k}{e^{\beta\epsilon_k}-1}
+\frac{\epsilon_k-(E_k+\mu)}{2}.
\eqname{Eex}
\end{equation}
An order by order comparison can be
performed in the high-temperature limit by expanding \eq{Esc} and
\eq{Eex} in powers of $\beta$: 
\begin{eqnarray}
\langle \mathcal{H}_1 \rangle_{SC}&\simeq& k_B
T-\frac{E_k+\mu}{2}+O\left[\beta(E_k+2\mu)^2\right] \\
\langle \mathcal{H}_1 \rangle_{\rm ex}&\simeq&
 k_B T-\frac{E_k+\mu}{2}+O(\beta\epsilon_k^2).
\end{eqnarray}
Agreement is found not only on the classical term $k_B T$, but also
on the subleading constant term $-(E_k+\mu)/2$, which would instead be missed
by a simple classical field theory. 

A more detailed comparison is obtained by working out
two limiting regions.
In the low energy limit, one has
\begin{eqnarray}
\lim_{\epsilon_k\rightarrow 0} \langle \mathcal{H}_1 \rangle_{\rm ex}&=& k_B T
-\frac{\mu}{2} \\
\lim_{\epsilon_k\rightarrow 0} \langle \mathcal{H}_1 \rangle_{\rm SC}&=& k_B T
-\frac{\mu}{2}+\frac{1}{6}\beta\mu^2+O(\beta^3\mu^4):
\end{eqnarray}
the relative error of the semi-classical result is therefore of the
order of $(\beta\mu)^2/6$, i.e. very small provided $k_B T \gg \mu$.

In the high energy limit $\epsilon_k\to\infty$, one has instead \cite{sim_math}
\begin{equation}
\langle \mathcal{H}_1 \rangle_{\rm SC} \sim
    \cosh(\beta\mu)\,\epsilon_k\,e^{-\beta\epsilon_k}.
\end{equation}
In the high temperature regime where $\cosh(\beta\mu)\simeq 1$,
this semi-classical prediction almost coincides
with the exact value \eq{Eex}
once the zero-point energy is subtracted from the quantum value. 
This shows that the semi-classical theory does not suffer from any 
ultraviolet divergence coming from the zero-point energy, nor from the
typical black-body catastrophe of classical field theories.

In summary, the semi-classical theory is able to accurately reproduce
the value of the average energy under the assumption that the
temperature is higher than the chemical potential, $k_B T \gg \mu$. 
Examples of plots of the mean energy of the different Bogoliubov modes
as a function of $E_k$ are presented in Fig.\ref{fig:Bogo_en} for the
semi-classical theory, the classical field approximation, and
the exact result.
The agreement of the semi-classical theory
with the exact result is already remarkable for
temperatures only a few times higher than the chemical potential, while
the classical field approximation is quite crude in predicting a constant
mean energy $k_B T$ independent of the mode energy.

Another observable that we consider is the {\em normal fraction} $f_n$,
defined as  
\begin{equation}
f_n=\frac{\langle P_x^2\rangle}{N m k_B T},
\label{eq:fn}
\end{equation}
where $P_x$ is the $x$ component of the total momentum of the system.
This quantity $f_n$ estimates the response of the Bose system to a gauge
field, e.g.\ a magnetic field in the case of charged particles, or a
rotation in the case of neutral ones~\cite{Leggett,Svistunov}.
 
The exact quantum result of the Bogoliubov theory \cite{attention} has
the form
\begin{equation}
\langle P_x^2\rangle_{\rm ex} = \sum_{\mathbf{k}\neq\mathbf{0}}
\hbar^2 k_x^2 n_k (n_k +1)
\end{equation}
where $n_k=(e^{\beta\epsilon_k}-1)^{-1}$ is the quantum mean occupation number of the
Bogoliubov mode.
The semi-classical approximation is instead given by
\begin{multline}
\langle P_x^2\rangle_{\rm SC} = 
\sum_{\mathbf{k}\neq\mathbf{0}}
\hbar^2 k_x^2 
\Big[\left(\sigma_R^2\right)_{\rm SC} \left(\sigma_I^2\right)_{\rm SC}
+\frac{1}{2} \left(\sigma_R^2\right)_{\rm SC} \\
+\frac{1}{2} \left(\sigma_I^2\right)_{\rm SC}
\Big].
\end{multline}

\begin{figure}[htbp]
\includegraphics[width=8cm,clip]{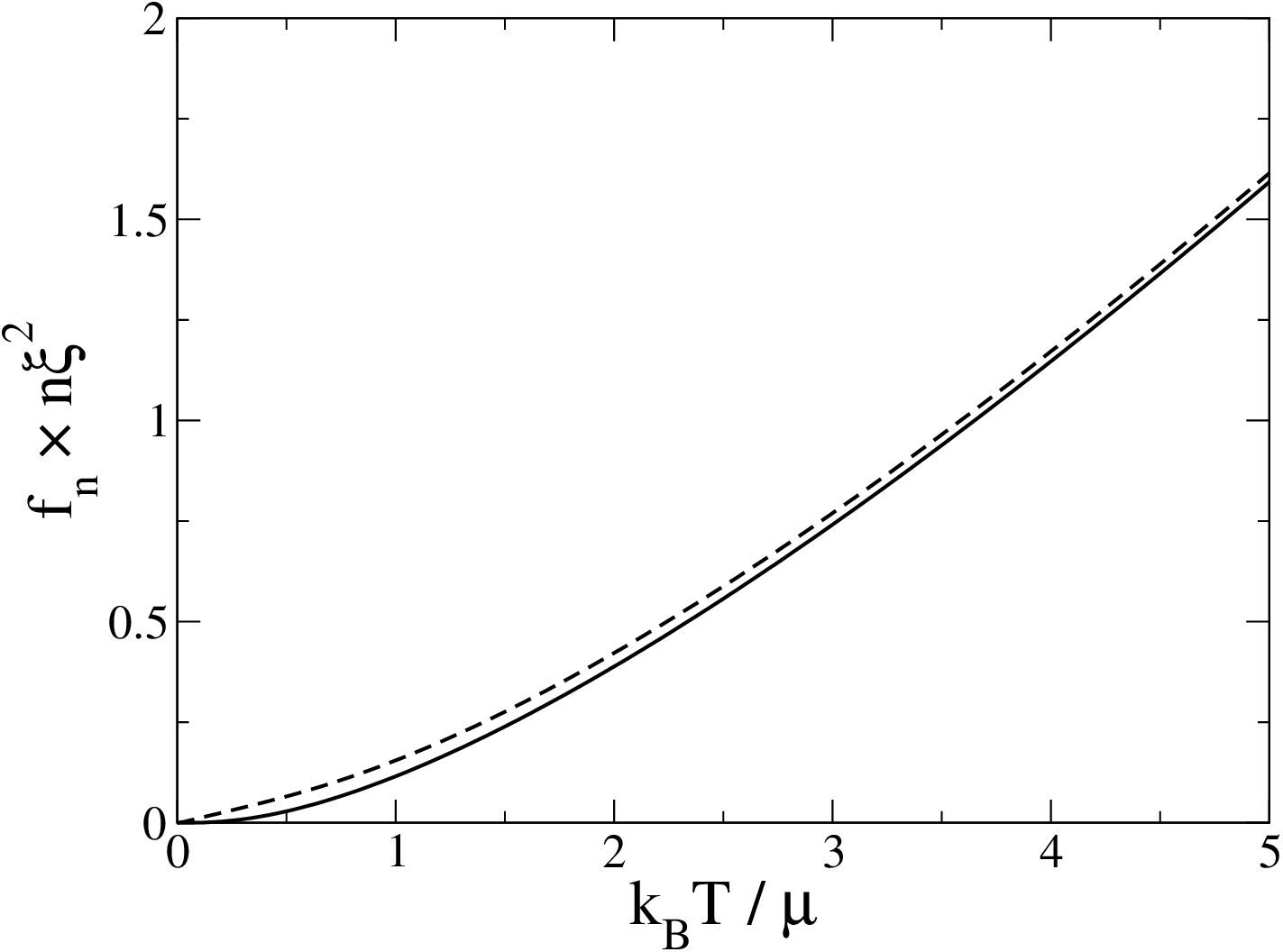}
\caption{For a two-dimensional
Bogoliubov gas in the thermodynamic limit,
normal fraction $f_n$ as a function of the temperature $k_B T$.
Solid line: quantum prediction. Dashed line: semi-classical prediction.
In order to have 
(within Bogoliubov theory) a universal function of $k_B T/\mu$, we
 actually plot the product of $f_n$ times $n\xi^2$, the healing   
length $\xi$ being defined by $\hbar^2/m\xi^2 = \mu$. }
\label{fig:fn}
\end{figure}

It is interesting to compare the expression between square brackets to
the quantum value $n_k (n_k+1)$,  at least in the high temperature
regime $k_B T \gg \mu$. 
For low momenta such that $E_k\leq \mu$, the semi-classical approximation
correctly reproduces the leading term $(k_B T/\epsilon_k)^2$ and
has an error $O(1)$.
The relative error is therefore of second order in $T$. 
For high momenta $\mu \ll E_k \simeq k_B T$, the semi-classical approximation
reproduces the quantum term with a relative error $O[(\beta \mu)^2]$.
After summation over all $\kk$ states, one finds for a two-dimensional 
Bogoliubov gas 
in the thermodynamic
limit that both the quantum and the semi-classical values of $f_n$ have
the form:
\begin{multline}
f_n = \frac{1}{2\pi n \xi^2}
\left\{\left[1+\ln\left(\frac{k_B T}{2\mu}\right)\right]\frac{k_B
 T}{\mu} \right. \\ \left.+\frac{1}{2} + O[\beta\mu\ln(\beta\mu)]\right\}, 
\end{multline}
where $\xi$ is the healing length defined by $\hbar^2/m\xi^2 = \mu$.
These results are summarized in Fig.\ref{fig:fn},
 where the semi-classical approximation for $f_n$ 
 is compared to the quantum value as a function of $k_B T/\mu$.

The last observable that we investigate is the
{\em pair distribution function},
\begin{equation}
g^{(2)}(\rr'-\rr) =\frac{1}{n^2}\,\Big\langle
\Psihd(\rr)\,\Psihd(\rr')\,\Psih(\rr')\,\Psih(\rr) \Big\rangle.
\label{eq:g2}
\end{equation}
Within the Bogoliubov approximation, this can be written for a two-dimensional system in the thermodynamic limit as: 
\begin{equation}
g^{(2)}(\rr)
\simeq 
1 + \frac{2}{n} \int \frac{d^2\mathbf{k}}{(2\pi)^2} \cos(\mathbf{k}\cdot\mathbf{r})
\left[\langle a_\mathbf{k}^\dagger a_\mathbf{k} + a_\mathbf{k} a_{-\mathbf{k}}\rangle\right]
\label{eq:g2Bog}
\end{equation}
where $n$ is the total density.
For a given $k$, the expectation value between square brackets in (\ref{eq:g2Bog})
is equal to $\sigma_R^2$.
Its value is given by Eq.(\ref{eq:srq}) for the quantum theory
and by Eq.(\ref{eq:srsc}) for the semi-classical theory.

\begin{figure}[htbp]
\includegraphics[width=8cm,clip]{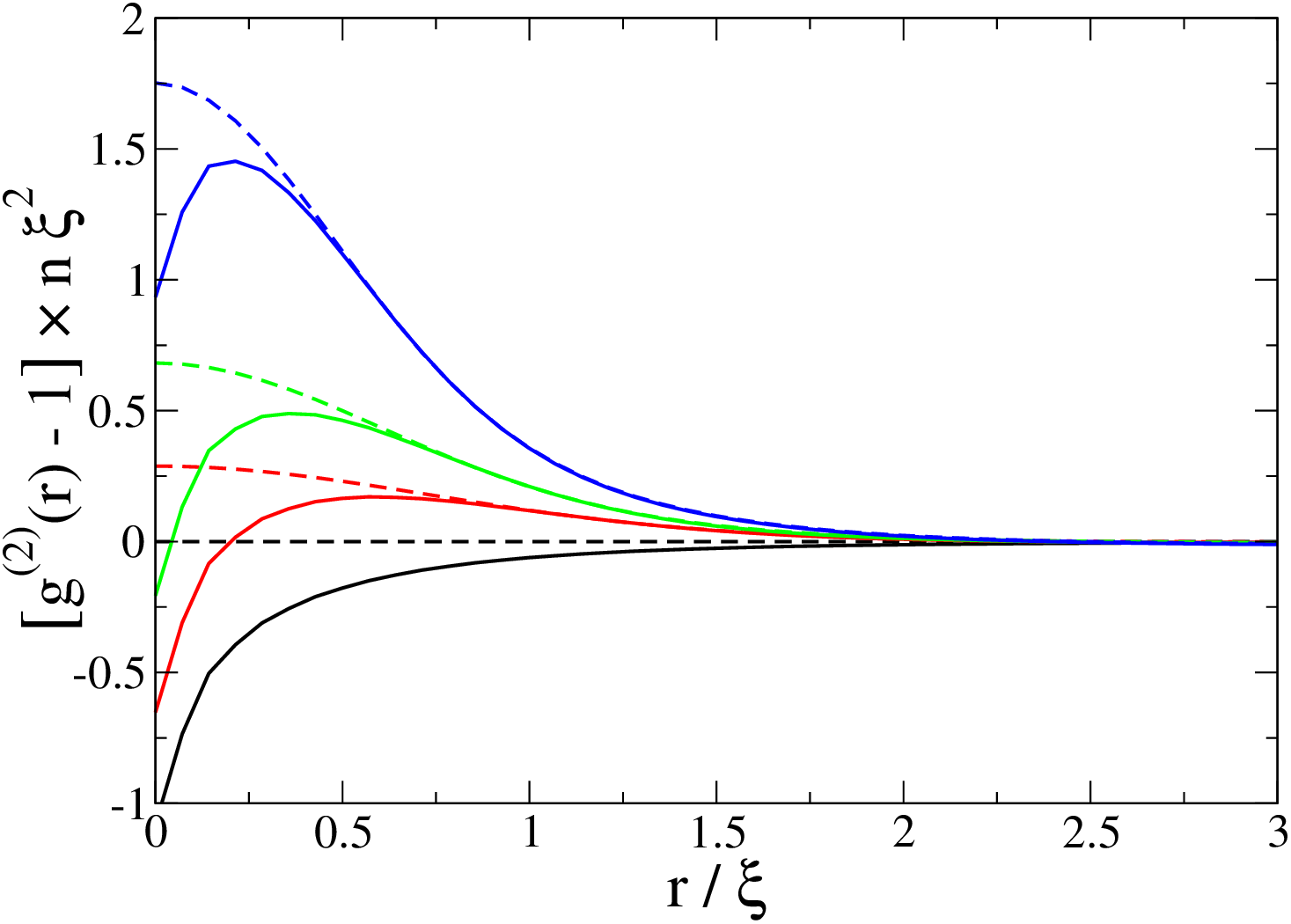}
\caption{(Color online) For a two-dimensional 
lattice Bogoliubov gas in the thermodynamic limit,
pair distribution $g^{(2)}(\mathbf{r})$ 
as a function of $r$ for different values
of the temperature, $k_B T /\mu = 0, 2, 3, 5$ (from bottom to top).
Solid line: quantum result. Dashed line: semi-classical approximation.
In the plot, the product of $g^{(2)}-1$ with  $n \xi^2$ is actually
 plotted, where $n$ is the density,  and $\xi$ the healing length such
 that  $\hbar^2/(m\xi^2)=\mu$. 
For the Bogoliubov gas, this product is indeed a universal function of
 $k_B T/\mu$ and $r/\xi$.
Here the lattice spacing is $0.07 \xi$.
 }
\label{fig:g2}
\end{figure}

In Fig.\ref{fig:g2} we plot the pair distribution $g^{(2)}(r)$ as a function
of $r$ for various values of the temperature.
The narrow dip which appears in the result of the quantum calculation
originates from the zero-point fluctuations of the Bogoliubov modes, and
is therefore absent in the semi-classical approximation:
in the quantum case, the decay of the Fourier transform of
$g^{(2)}(\mathbf{r})-1$ at large $k$ is in fact algebraic, whereas it is
Gaussian in the semi-classical approximation.
On the other hand, the semi-classical approximation
reproduces remarkably well the intermediate to long-distance
behavior already at temperatures as low as $k_B T = 2 \mu$. 

\subsection{In the canonical ensemble}
\label{subsec:itce}

In the language of~\cite{IYJ_QMC}, the semi-classical method discussed in
the previous sections can be seen as a ``simple coherent'' scheme from 
which the noise terms have been dropped. 
This suggests that a similar procedure may be applied to the ``simple
Fock'' scheme in order to devise a semi-classical method that works in
the canonical ensemble, i.e. at a fixed number $N$ of particles.

The building block of this scheme is the Fock state defined as usual as:
\begin{equation}
|N:\psi\rangle=\frac{1}{\sqrt{N!}}\,(\ahd_\psi)^N\,|0\rangle,
\end{equation}
$|0\rangle$ is here the vacuum state and the $\ahd_\psi$ operator
creates a particle in the (not necessarily normalized) $\psi$ state:
\begin{equation}
\ahd_\psi=\sum_\rr dV\,\psi(\rr)\,\Psihd(\rr).
\end{equation}
By projecting both sides of \eq{exp_coh} onto the subspace with exactly 
$N$ particles,
it is easy to see that any $N$-body density operator can be expanded on 
a family of Fock states as:
\begin{equation}
\rho=\int_{||\psi||=1}\!{\mathcal D}\psi\,\mathcal{P}[\psi]
\,|N:\psi\rangle\langle
 N:\psi|,
\eqname{ansatz_Fock}
\end{equation}
where the distribution $\mathcal{P}$ is the 
Fock state equivalent of the Glauber-P distribution, and 
the integral is taken over the unit sphere $||\psi||=1$.
The infinite temperature state $\rho_{\rm eq}(\tau=0)={\mathbf 1}$ is
obtained by simply taking a constant value for $\mathcal{P}[\psi]$.
This corresponds to a random selection of the wavefunction
$\psi(\tau=0)$ with a uniform distribution on the unit sphere
$||\psi||=1$. 
At finite temperature, the distribution function $\mathcal{P}[\psi]$ for an
interacting gas is unfortunately not necessarily regular and positive; 
as a consequence, no stochastic evolution for $\psi$ 
exists such that the thermal density operator $\rho(\beta)$ is obtained
as the average of dyadics of the form $|N:\psi\rangle\langle N:\psi|$.
On the other hand, one can find a stochastic evolution ensuring that  
$\rho(\beta)$ is the average of dyadics of the slightly different form
$|N:\psi_1\rangle\langle N:\psi_2|$.
$\psi_1$ and $\psi_2$ are here independent realizations of the Ito
stochastic process~\cite{IYJ_QMC}
\begin{multline}
d\psi(\rr) =-\frac{d\tau}{2}\left[h_0+
g_0\frac{N-1}{\| \psi \|^2}\,|\psi(\rr)|^2 \right. \\
\left. - g_0\frac{N-1}{2}\frac{\sum_{\rr'} dV\,|\psi(\rr')|^4}{\|\psi\|^4} 
\right]\,\psi(\rr) +dB(\rr),
\eqname{evol_Fock}
\end{multline}
starting from the common value $\psi(\tau=0)$, and the
correlation functions of the noise $dB(\rr)$ satisfy the condition:
\begin{equation}
dB(\rr)\,dB(\rr')=-\frac{g_0 d\tau}{2 dV} \mathcal{Q}_\rr
\mathcal{Q}_{\rr'} \left[\delta_{\rr,\rr'}\psi(\rr) \psi(\rr')\right],
\end{equation}
where the projector $\mathcal{Q}$ projects orthogonally to the ket
$|\psi\rangle$.

 From this exact reformulation of the full many-body problem, it is
immediate to obtain a canonical version of the semi-classical scheme by
simply neglecting the noise term $dB$ in \eq{evol_Fock}.
Intuitively this is expected to constitute a good approximation of the 
quantum model at least in the high-temperature case, i.e.\ for 
`times' $\tau$ short enough for the effect of the noise terms to remain small.
The corresponding semi-classical approximation of
the density operator for the thermal equilibrium state
at temperature $T$ in the canonical ensemble is therefore
\begin{equation}
\rho_{\rm SC} = \int_{||\psi(0)||=1} \mathcal{D}\psi(0) \,
|N:\psi(\beta)\rangle \langle N:\psi(\beta)|,
\eqname{rho_SC}
\end{equation}
where $\psi(\beta)$ has evolved from its initial value $\psi(0)$
during a `time' $\beta$
according to the deterministic part of \eq{evol_Fock}, 
\begin{multline}
\partial_\tau\psi(\rr,\tau) =-\frac{1}{2}\left[h_0+
g_0\frac{N-1}{\| \psi \|^2}\,|\psi(\rr,\tau)|^2 \right. \\
\left. - g_0\frac{N-1}{2}\frac{\sum_{\rr'} dV\,|\psi(\rr',\tau)|^4}{\|\psi\|^4} 
\right]\,\psi(\rr,\tau),
\label{eq:sogpe}
\end{multline}
which closely ressembles an imaginary time Gross-Pitaevskii equation.

This semi-classical Fock scheme can be used as the core of a
numerical Monte Carlo code to study the 
properties of a $N$-body Bose gas at thermal equilibrium. 
 From the computational point of view, the only non trivial aspect is how
to efficiently perform the sampling of $\psi(0)$ on the unit sphere.
The numerical algorithm that we have adopted for this purpose is detailed
in the appendix \ref{appen:nace}. 

\section{Application to thermal vortices in the 2D gas}
\label{sec:appl}
In this second part of the paper, we apply the semi-classical technique
developed in the first part to the study of some among the most
significant properties of a homogeneous two-dimensional Bose gas at
thermal equilibrium in the canonical ensemble.
This problem of the 2D Bose gas is under active experimental investigation.
It is known theoretically that the 2D Bose gas exhibits the Berezinskii-Kosterlitz-Thouless
transition \cite{BKT,Minnhagen,Markus}, and this transition was recently observed
with cold atoms in \cite{Dalibard2D}.
An interesting aspect of the experiments with atoms is that they have
access to vortices \cite{Dalibard2D,Cornell_APS}, so that
special attention will be paid here
to observables such as the density and the
pair distribution function of thermally activated
vortices, for which classical field methods \cite{Hutch}
and in particular the present semi-classical field method, are
well suited.
Our numerical results will then be interpreted in terms of simplified
analytical models, which allow one to unravel the underlying physics.

The model Hamiltonian used to describe the system is the
two-dimensional version of the spatially homogeneous lattice model
\eq{Hamilt} with periodic boundary conditions. 
The value of the coupling constant $g_0$ to be used in the calculations
depends on the details of the atomic confinement along the third
dimension: here, we assume a harmonic confinement in the $z$ direction,
with a harmonic oscillator length $a_{\rm ho} = \sqrt{\hbar/m\omega_z}$  
much larger than the three-dimensional $s$-wave scattering length
$a_{\rm 3D}$.  
In this limit, one is allowed to neglect the energy-dependence of the
effective two-dimensional coupling constant $g$
\cite{ShlyapHouches,note}, and to simply take \cite{Dum2D} 
\begin{equation}
g_0 = \frac{\hbar^2}{m} \frac{2\sqrt{2\pi} a_{\rm 3D}}{a_{\rm ho}}.
\end{equation}
To ensure the two dimensional character of the atomic gas, we assume 
that both the thermal energy $k_B T$ and the mean field
zero-temperature chemical potential $g_0 n$ are much smaller than the
confinement energy $\hbar\omega_z$ in the $z$ direction.
Note that the semi-classical approach is limited to the weakly
interacting gas regime $n \xi^2 \gg 1$, the healing length $\xi$
being defined by $\hbar^2/m \xi^2 = n g_0$. Remarkably, this condition
reduces to the density-independent one $m g_0/\hbar^2 \ll 1$ in two
dimensions.  

\subsection{Numerical results}
\label{sec:numerical}
\subsubsection{Normal and non-condensed fractions}
\begin{figure}[h]
\vspace{1cm}
\includegraphics[width=8cm,clip]{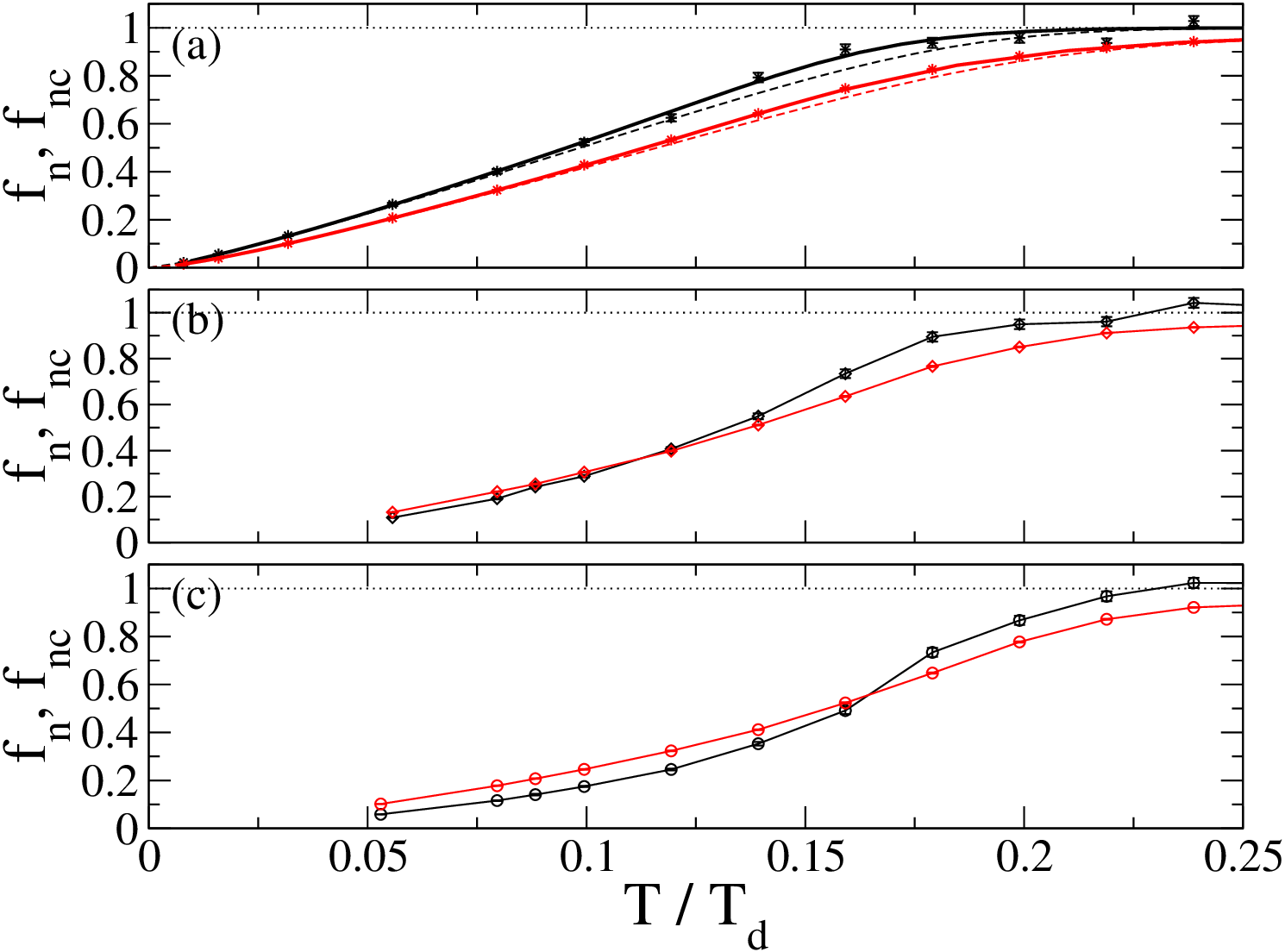}
\caption{(Color online) Normal fraction $f_n$ (black) and non-condensed fraction
 $f_{\rm nc}$ (red) as functions of temperature for a two-dimensional
 Bose gas with $N=1000$ particles on a square box of size $L$ with
 periodic boundary conditions.
(a) Ideal Bose gas. 
(b) Interacting gas with a coupling constant $g_0=0.1 \hbar^2/m$.
(c) Interacting gas with $g_0=0.333 \hbar^2/m$.
Symbols: results of semi-classical simulations on a 64$\times$64 grid with 2000 realizations.
Solid lines: in (a) exact result from the canonization procedure (see text); in (b) and (c),
a guide to the eye.
Dashed lines in (a): the grand canonical predictions.
The temperature is in units of the degeneracy temperature $T_d$ such that $k_B T_d=2\pi \hbar^2 n/m$.}
\label{fig:fnfnc}
\end{figure}

The normal fraction \eq{fn} describes the response of the fluid to
a spatial twist of the phase~\cite{Leggett,Svistunov}, while 
the non-condensed fraction is simply the fraction of atoms in
single-particle states other than the zero-momentum plane wave
$f_{\rm nc}=1-N_0/N$.
These two quantities are plotted in Fig.\ref{fig:fnfnc} as functions 
of the temperature for three different values of the interaction
strength $g_0$, including the ideal gas $g_0=0$.
The overall behavior is almost the same for all the curves: the
dependence on temperature is always smooth and, as expected, both the
normal and the non-condensed fractions tend to $1$ ($0$) in respectively
the high (low) temperature limit.
However, whereas the shape of the curve giving the non-condensed
fraction is not qualitatively modified as $g_0$ grows, the crossover
from $0$ to $1$ of the normal fraction turns out to become somehow
sharper as the interaction strength is increased~\cite{pas_thermo}. 

It is interesting to compare the results for the ideal gas case with a
(trivial) calculation performed in the grand canonical ensemble: as one
can see in Fig.\ref{fig:fnfnc}a, the dashed line corresponding to the
grand canonical prediction significantly deviates from the numerical
simulation results.
A simple explanation for this can be put forward in terms of the finite
size of the system, which can indeed lead to differences between the two
ensembles. In particular for a Bose condensed ideal gas,
the grand canonical ensemble predicts
unphysically large fluctuations of the number of condensate particles
\cite{Wilkens,Olshanii,Holthaus};
although this does not significantly affect the normal and
the non-condensed fractions plotted here, 
it  will have a dramatic impact on other quantities
like the density fluctuations and the mean vortex density that will be 
studied in what follows.

In order to fully clarify this issue, an exact canonical calculation can 
be performed by means of the standard canonization procedure
\cite{canon}: the analytical predictions for the normal and the 
non-condensed fractions are plotted in Fig.\ref{fig:fnfnc}a and compared
to the Monte Carlo ones. The agreement is remarkable.

\subsubsection{Density fluctuations} 
\begin{figure}[h]
\vspace{1cm}
\includegraphics[width=8cm,clip]{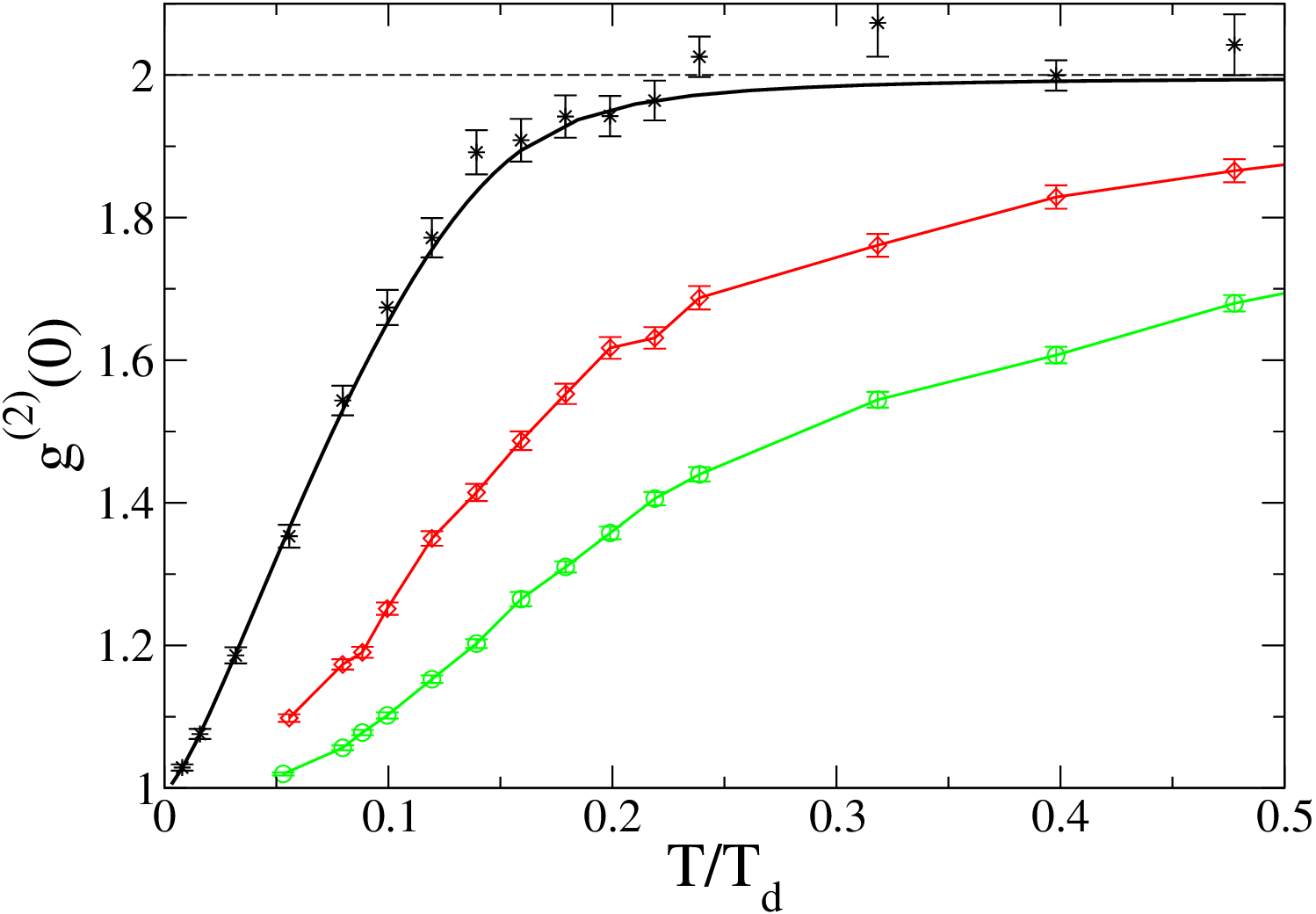}
\caption{(Color online) Pair distribution function $g^{(2)}(0)$ as a function of temperature
for the same parameters as in Fig.\ref{fig:fnfnc}.
Symbols: results of the semi-classical simulations.
 From top to bottom, the value of the coupling constant increases from
 $g_0=0$ (black stars) to $g_0=0.1\hbar^2/m$ (red diamonds) and $0.333
 \hbar^2/m$ (green circles). 
Solid lines: for $g_0=0$ the exact result from the canonization
 procedure, for $g_0>0$ a guide to the eye.
Horizontal dashed line: grand canonical prediction $g^{(2)}(0)=2$ for
the ideal gas. 
The temperature is in units of the degeneracy temperature $T_d$ such that $k_B T_d=2\pi \hbar^2 n/m$.}
\label{fig:g2sc}
\end{figure}

In Fig.\ref{fig:g2sc} we plot the temperature dependence of the pair
distribution function \eq{g2} of the gas evaluated at coincident
points $\rr=\rr'$, i.e. $g^{(2)}(0)$ \cite{expm}.
In~\cite{Svistunov_tc2db} this quantity was related in a classical field
model to
the notion of a quasi-condensate density in the low temperature superfluid
regime, $n_{QC}=n\,\sqrt{2-g^{(2)}(0)}$.
In the figure, the dependence of $g^{(2)}(0)$ is shown for three
values of the interaction strength $mg_0/\hbar^2=0,\, 0.1,\,0.333$. 
In the ideal gas case $g_0=0$, the Monte Carlo results are in remarkable agreement 
with the exact canonical results obtained from the canonization procedure ~\cite{thermo2};
on the other hand,
at low temperatures, 
when a significant condensed fraction is present,
the grand canonical prediction $g^{(2)}(0)=2$ strongly differs
from the canonical results and becomes physically incorrect.
Concerning the dependence on the interaction strength $g_0$, our
simulations confirm the expected trend that an increase of the interaction strength $g_0$
at a fixed value of the non-condensed fraction corresponds to a strong decrease of the density
fluctuations.

Comparing Fig.\ref{fig:g2sc} to Fig.\ref{fig:fnfnc}, it is immediate to
see that density fluctuations are already significant in the range of
temperatures corresponding to the rapid increase of the normal fraction.
This shows that density fluctuations may play an important role
in the superfluid transition of a 2D gas~\cite{Safonov,Svistunov2D}.

\subsubsection{Vortex density}
In the semi-classical theory, it is straightforward to define a vortex
density by looking for the vortices that appear in each stochastic
realization of the classical field $\psi(\rr)$. This is an
advantage with respect to e.g.\ Path Integral Quantum Monte Carlo
methods~\cite{QMC}.

The field $\psi(\rr)$ of the semi-classical method, initially defined on
a lattice, may be extended to any point of the continuous space by means
of the Fourier formula  
\begin{equation}
\psi(\rr) = \frac{1}{L} \sum_{\kk} a_\kk e^{i\kk\cdot\rr},
\end{equation}
where the $a_\kk$ are the Fourier components of the field on the
lattice.
As usual, vortices correspond to nodes in the field $\psi$ with a
non-zero circulation; numerically, they can be efficiently and precisely
located by calculating the circulation of the phase gradient of the
field $\psi$ around plaquettes of much smaller size than the original
lattice cell \cite{ea}.

\begin{figure}[htbp]
\includegraphics[width=8cm,clip]{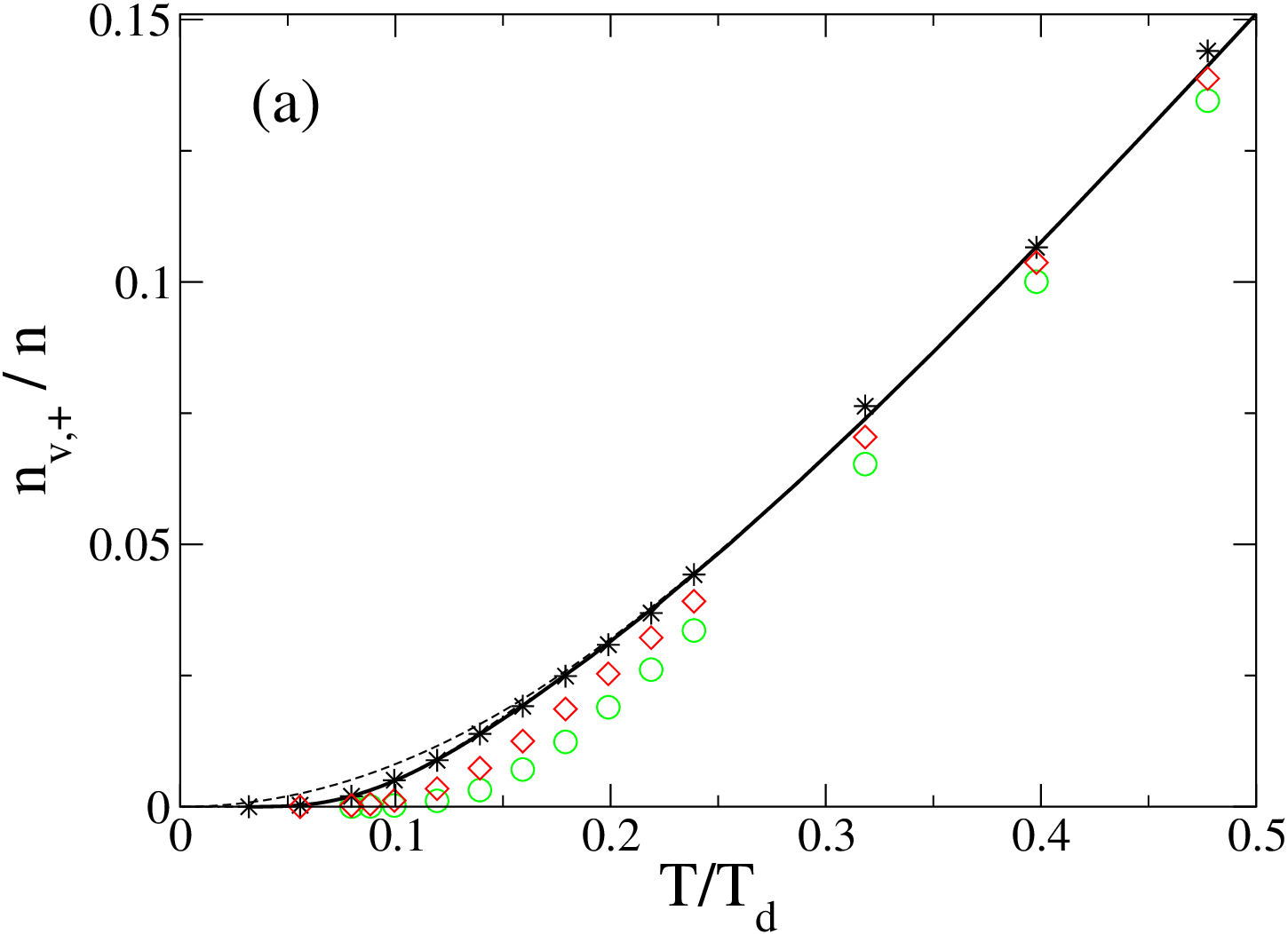}
\vspace{1cm}
\includegraphics[width=8cm,clip]{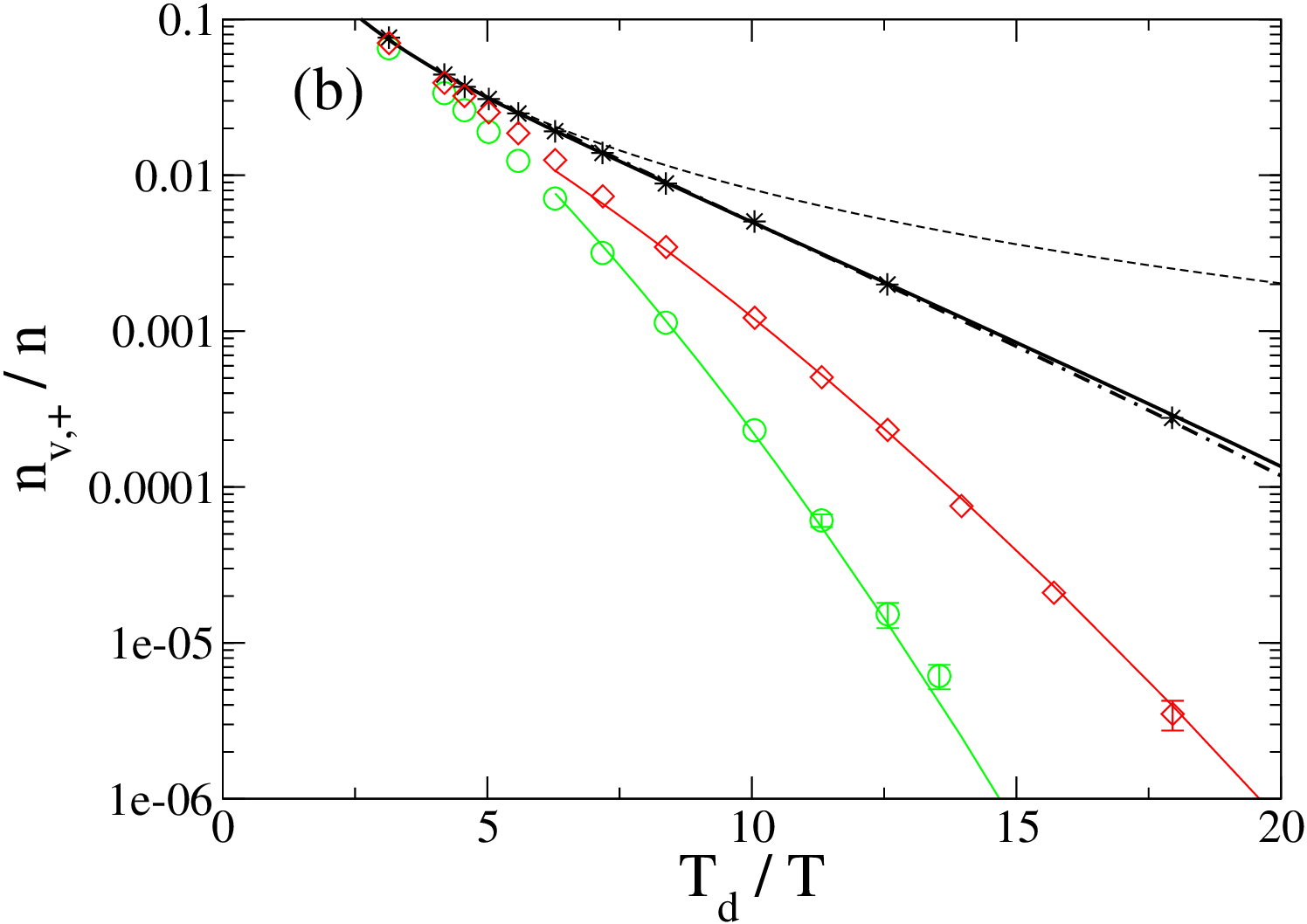}
\caption{(Color online)
Mean density of positive charge vortices as a function of temperature for various
 interaction strengths. 
The parameters have the same values as in Fig.\ref{fig:fnfnc}. 
(a) Linear scale, (b) logarithmic scale for the vortex density, reciprocal 
 scale for the temperature.
Symbols: results of the semi-classical simulation, $g_0=0$ (black
 stars), $g_0=0.1\hbar^2/m$ (red diamonds), 
$g_0=0.333 \hbar^2/m$ (green circles).
 Solid lines : the exact canonical result \eq{nv+can} for $g_0=0$;
 prediction of the activation law model of Sec.\ref{sec:analyt} for
 $g_0>0$, $n_{v,+}/n=C e^{-\Delta(T)/k_B T}$, with the 
 prefactor $C$ taken as a constant and fitted to the data
 ($C=0.134$ for $g_0=0.1\hbar^2/m$ and
 $C=0.3355$ for $g_0=0.333\hbar^2/m$).
Dashed line: grand canonical result for $g_0=0$.
Dot-dashed line: Bogoliubov prediction for $g_0=0$ for $T/T_d<0.15$,
essentially indistinguishable from the solid line in (a).
Note that the circle with the largest value of $T_d/T$
corresponds to $k_B T/n g_0\simeq 1.4$, 
which is on the limit of the validity
of both the semi-classical field method and of the simple model 
of section \ref{sec:analyt} calculating $\Delta$.
\label{fig:nv}
}
\end{figure}

Numerical results for the mean density of positive charge vortices
$n_{v,+}$ as a function of temperature for various interaction strengths
are shown in Fig.\ref{fig:nv}a. 
Thanks to the periodic boundary conditions, each
realization of the field has the same number of positively and
negatively charged vortices, which implies $n_{v,-}=n_{v,+}$. 
For the considered finite size system, there is no qualitative
difference between an ideal and an interacting gas: in both cases, the
vortex density varies roughly linearly with temperature at high
temperature, while it decreases very rapidly at low
temperature. 
Looking at the same data on the logarithmic-reciprocal scale of
panel (b), it is easy to observe that the low temperature decrease
of  $n_{v,+}$ roughly follows an activation law of the form $\propto   
e^{-\Delta /k_B T}$.
A thorough and analytic explanation of this central issue will be
given in section \ref{subsec:Bog} for the non-interacting $g_0=0$ case and
in Sec.\ref{sec:analyt} for the general case.

\subsubsection{Pair distribution function for vortices} 
\begin{figure}[htbp]
\includegraphics[width=8cm,clip]{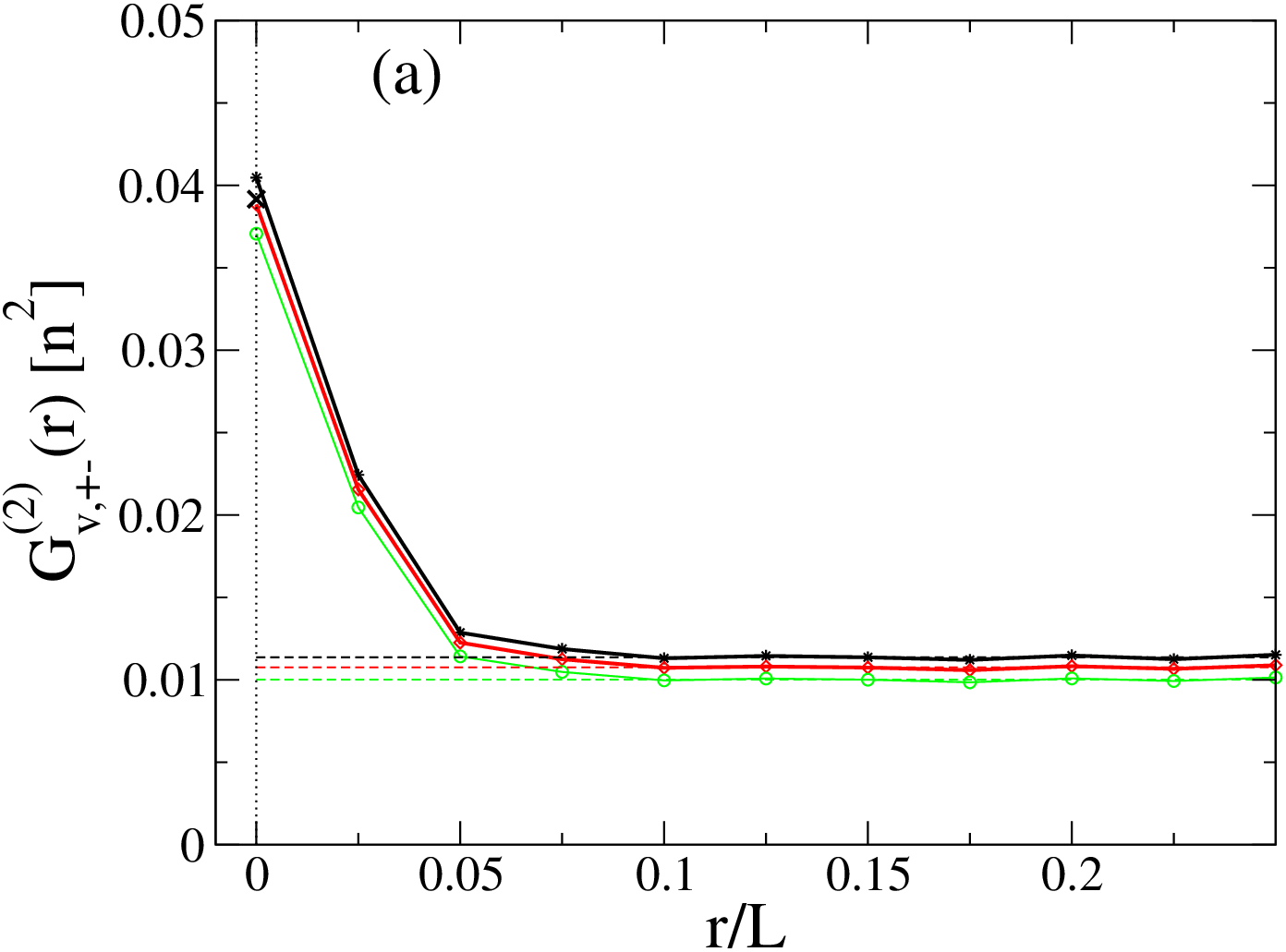}
\vspace{0.5cm}
\includegraphics[width=8cm,clip]{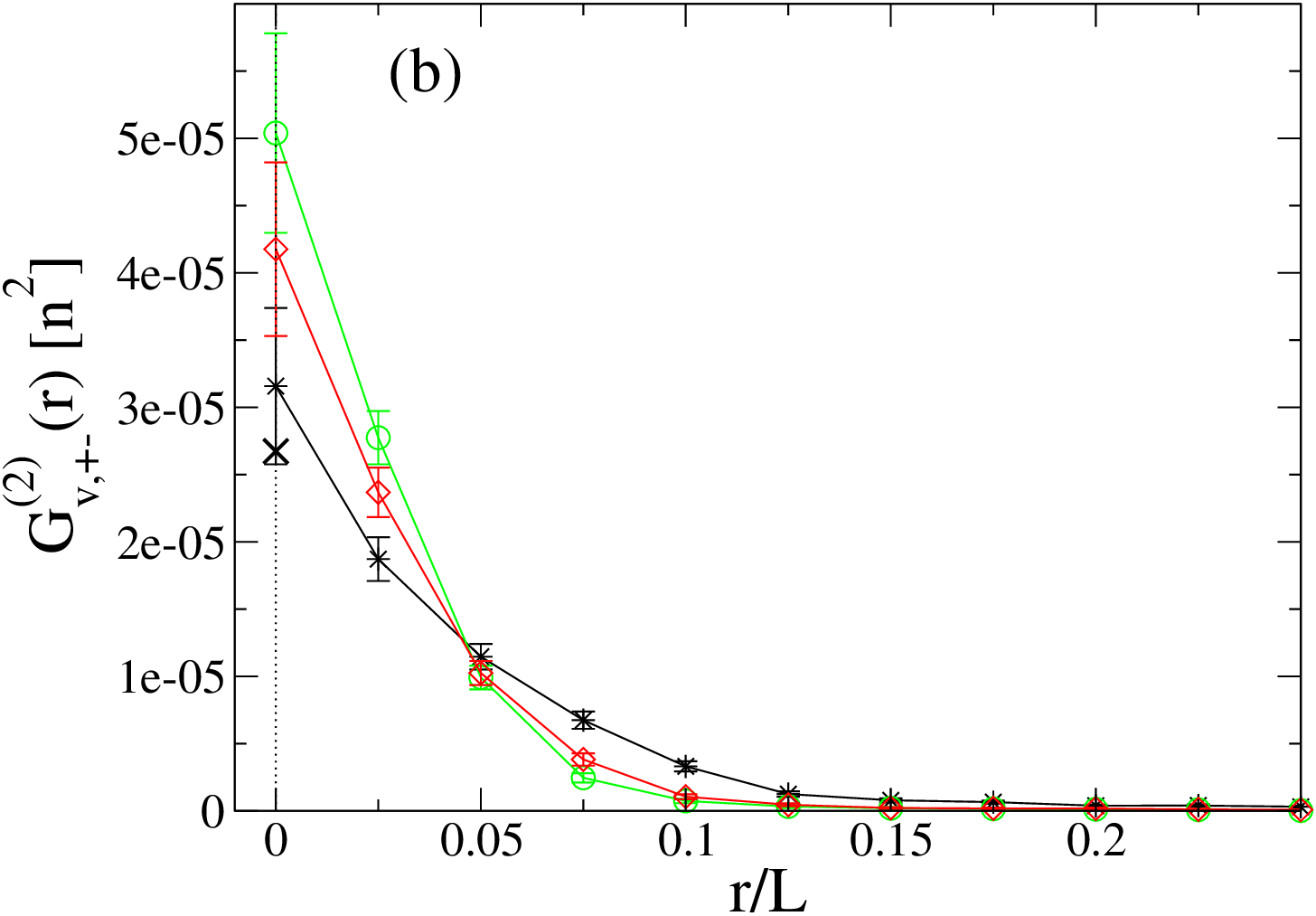}
\caption{(Color online)
Results of the semi-classical simulations for the angular average
$G^{(2)}_{v,+-}(r)$ of the pair distribution function 
for opposite charge vortices as a function of the
distance $r$ between the two vortices. 
The parameters have the same values as in Fig.\ref{fig:fnfnc}.
(a) High-temperature, non-Bose condensed regime, temperature 
$T/T_d=2.5/(2\pi)\simeq 0.398$, for $m
 g_0/\hbar^2=0$ (black stars), $0.1$  (red diamonds), $0.333$ (green
 circles).  
The solid lines are a guide to the eye.
Horizontal dashed lines: square of the mean vortex density
$n_{v,+}^2$, showing the decorrelation at long distances. 
(b) Low temperature, Bose-condensed regime. The temperatures are adjusted
 to have similar vortex densities for the various values of $g_0=0$
 (black stars, $T/T_d=0.35/(2\pi)\simeq 0.056$, leading to $n_{v,+} \simeq
 0.28/L^2$), 
 $g_0=0.1 \hbar^2/m$ (red diamonds, $T/T_d=0.5/(2\pi)\simeq 0.08$, leading
 to 
 $n_{v,+}\simeq 0.23/L^2$), $g_0=0.333 \hbar^2/m$ (green circles,
 $T/T_d=0.625/(2\pi)\simeq 0.1$, leading to $n_{v,+}\simeq 0.23/L^2$).
The solid lines are a guide to the eye.
In both panels (a) and (b), the cross at $r=0$ gives the exact value of
 $G^{(2)}_{v,+-}$ for the ideal gas, obtained with the canonization
 procedure. 
The distance $r$ is in units of $L$ and $G^{(2)}_{v,+-}$ is in units of
the squared particle density $n^2$.
}
\label{fig:G2v}
\end{figure}

As a last observable, it is interesting to look at the pair distribution
function for vortices. 
In analogy with the pair distribution functions for particles in a gas, 
and restricting for simplicity our attention to the case of opposite
charge vortices, this may be defined as
\begin{equation}
G^{(2)}_{v,+-}(\rr) = \langle \rho_{v,+}(\mathbf{0}) \rho_{v,-}(\rr)\rangle.
\end{equation}
For a given realization of the field, $\rho_{v,\pm}(\rr)$ is
here the sum of Dirac deltas $\delta(\rr-\rr_{v,\pm})$ centered on the
locations $\rr_{v,\pm}$ of the positive (respectively negative) charge
vortices.
The angular average of $G^{(2)}_{v,+-}$ is plotted as a function of the
distance $r$ in Fig.\ref{fig:G2v} for different values of the coupling
constant $g_0$ and temperature.

In Fig.\ref{fig:G2v}a, a high temperature (but still degenerate) case is
considered, where both the normal and the non-condensed fractions are
close to unity: a peak appears in all curves at $r=0$ as well as a
plateau at larger vortex separations $r$.
The former is a consequence of the effective attraction among opposite
charge vortices, while the latter corresponds to the decorrelated value
$G^{(2)}_{v,+-}\simeq n_{v,+} n_{v,-}$.
These numerical results indicate a weak dependence on the interaction
strength, and are in good agreement with the known result (not shown) 
for the ideal gas in the grand canonical ensemble~\cite{Halperin,Berry,noteG2V}.
 
In Fig.\ref{fig:G2v}b, the considered temperatures are
low enough to be in the regime where $n_{v,+}$ drops very
rapidly with $T$. 
For each value of the interaction strength $g_0$, the temperature is
selected to give a roughly fixed vortex density.
A noticeable difference between the ideal and the interacting gas
cases appears: the correlations between opposite charge vortices
have a much longer range in the ideal gas than in the interacting one.

\begin{figure}[htbp]
\includegraphics[width=6cm,clip]{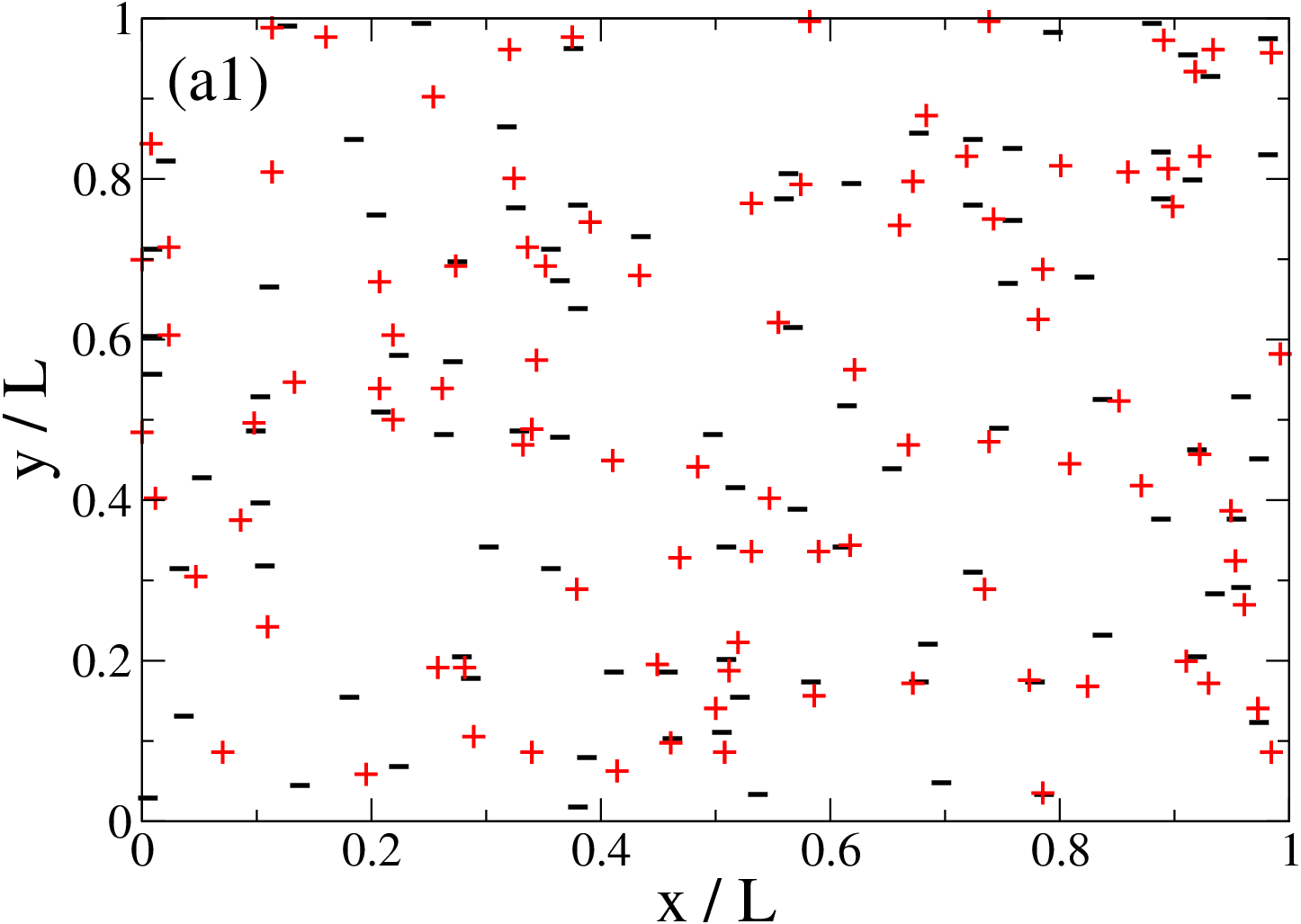}
\vspace{0.5cm}
\includegraphics[width=6cm,clip]{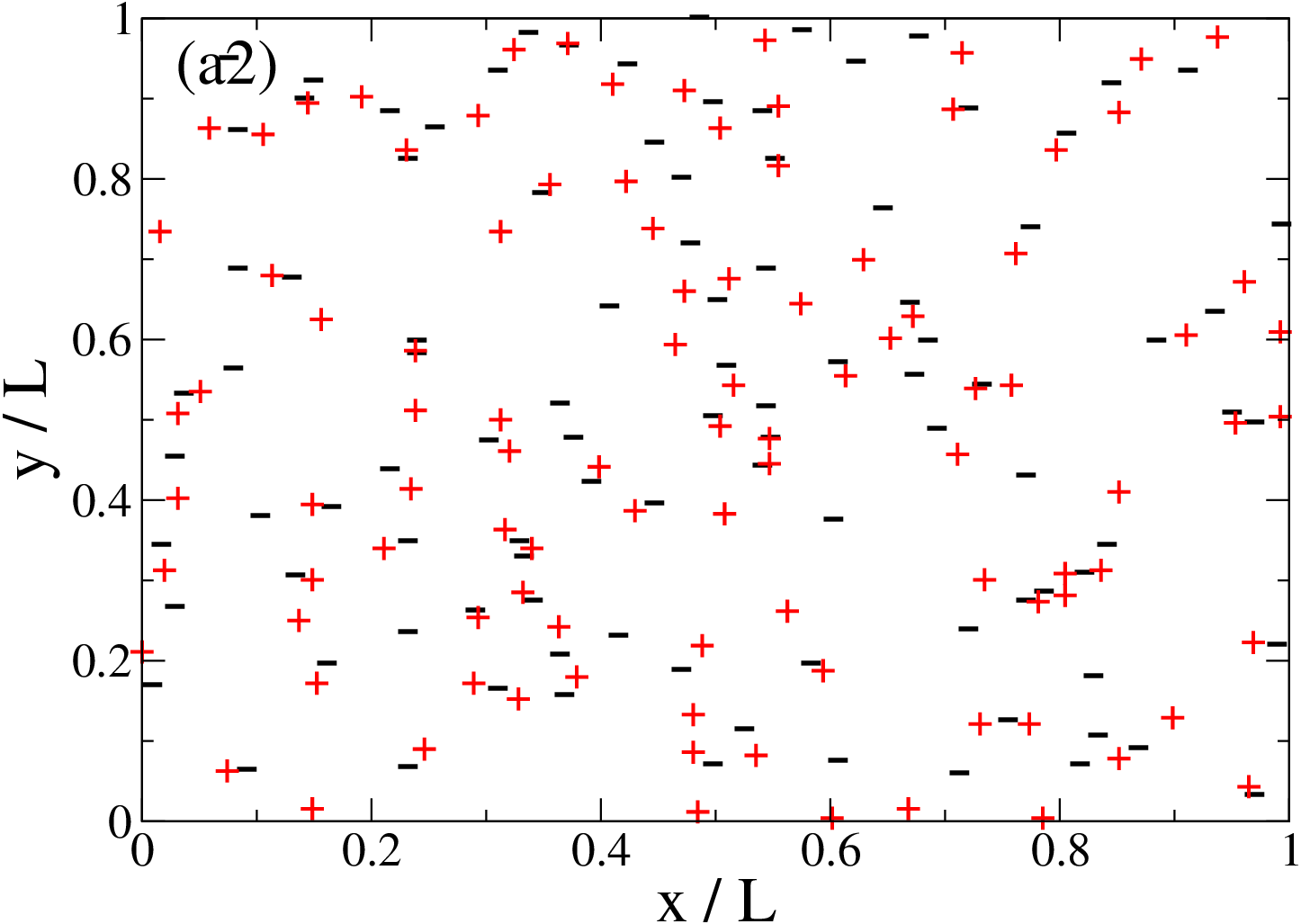}
\vspace{0.5cm}
\includegraphics[width=6cm,clip]{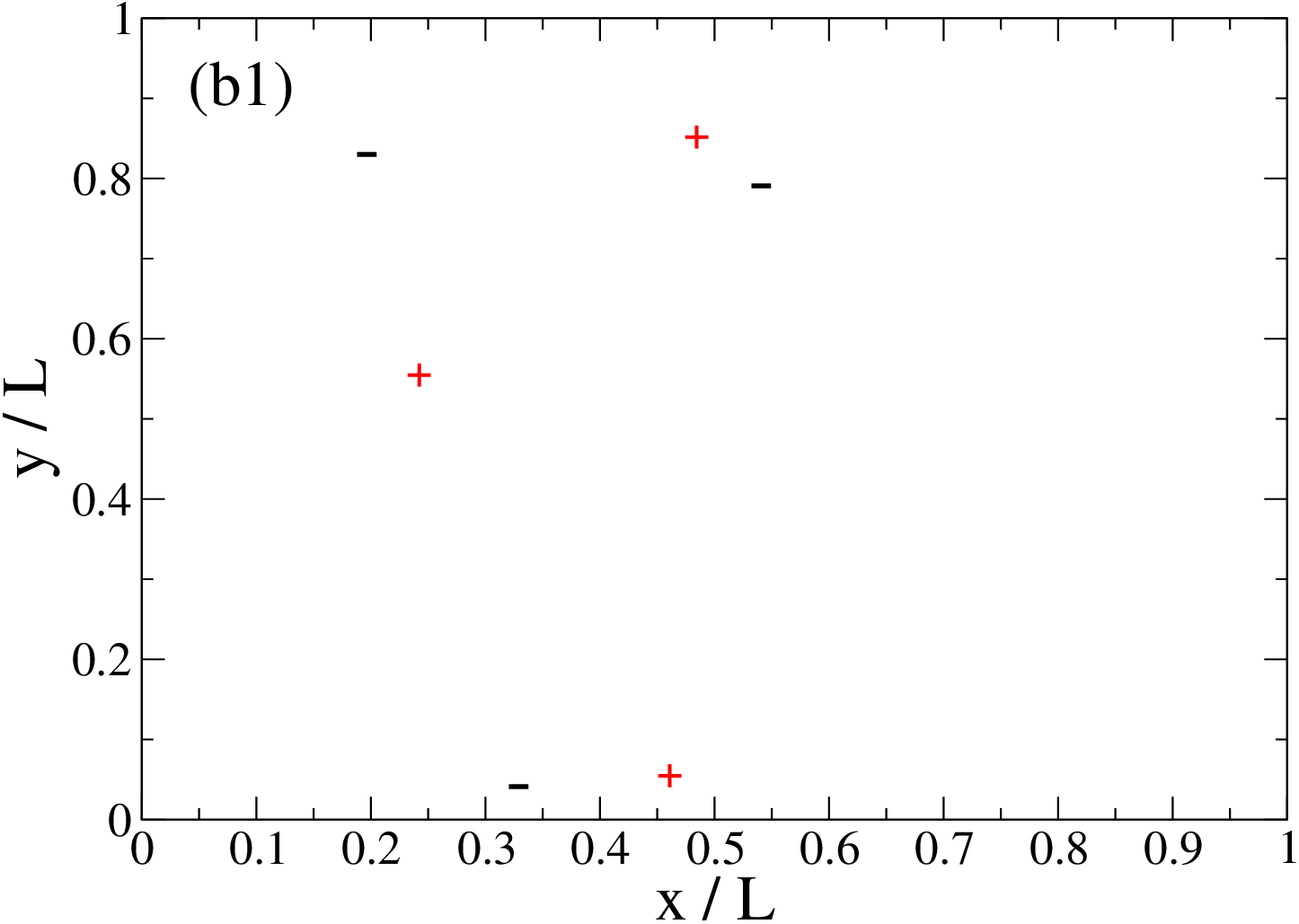}
\vspace{0.5cm}
\includegraphics[width=6cm,clip]{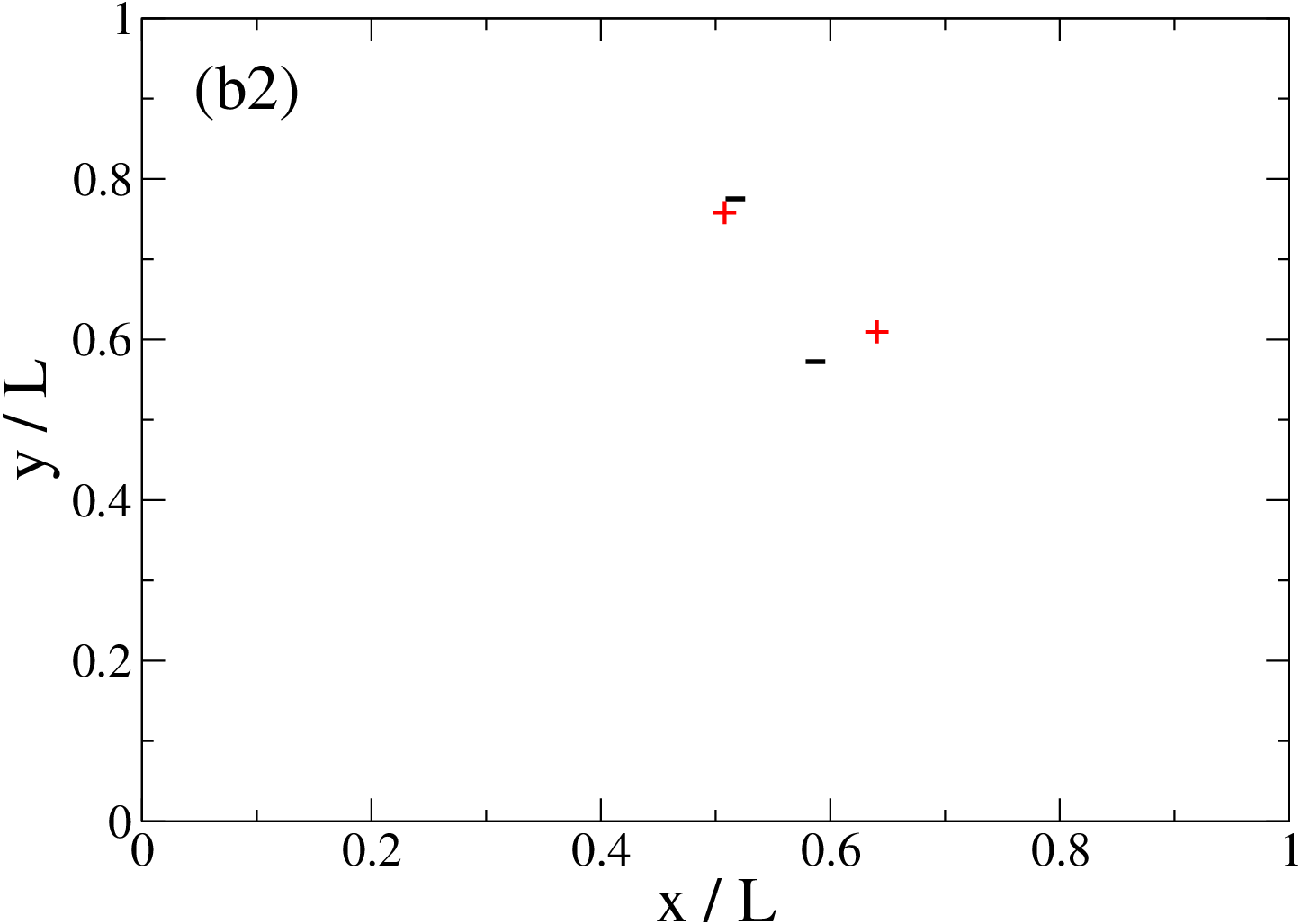}
\caption{(Color online)
For arbitrary Monte Carlo realizations of the field with vortices, locations of the
positive charge vortices (red plus symbols) and negative charge vortices
(black minus symbols) in the field.
Parameters as in some curves of Fig.\ref{fig:G2v}:
(a1) $T/T_d=2.5/(2\pi)\simeq 0.398$ for $g_0=0$.
(a2) $T/T_d=2.5/(2\pi)\simeq 0.398$ for $g_0=0.333 \hbar^2/m$.
(b1) $T/T_d=0.35/(2\pi)\simeq 0.056$ for $g_0=0$.
(b2) $T/T_d=0.625/(2\pi)\simeq 0.1$ for $g_0=0.333 \hbar^2/m$.
Note that the realizations shown in 
(b1) and (b2) are not fully typical since they contain several pairs.
\label{fig:visual}}
\end{figure}

A more intuitive representation of these issues is given in
Fig.\ref{fig:visual}, where the locations of the vortices are shown for
some randomly selected Monte Carlo realizations of the field. 
The high temperature case is considered in (a1) for the ideal
gas and in (a2) for the interacting gas.
The effect of interactions in the low-temperature regime is visible 
in panels (b1) and (b2): the difference in behavior between the ideal
(b1) and the interacting (b2) gas cases is apparent, the vortex pairs in
the ideal gas being much larger.

\subsection{The effect of Bose condensation on the vortex density in an
   ideal gas: Bogoliubov theory}
\label{subsec:Bog}

To understand the simulation results for the vortex density in the
non-interacting case, a naive approach is to use the grand
canonical ensemble. In this case,
the Glauber-P distribution for the field
is indeed Gaussian, so that exact analytical predictions 
can be obtained for the vortex density \cite{Halperin,Berry}:
\begin{equation}
\left(n_{v,+}\right)_{\rm GC}= \frac{m}{4\pi \hbar^2} \frac{\sum_\kk E_k n_k}
{\sum_\kk n_k},
\end{equation}
where $E_k=\hbar^2 k^2/2m$, the mean occupation numbers are given by the
Bose formula, $n_k=1/\{\exp[\beta (E_k-\mu)]-1\}$, and the chemical
potential $\mu$ is adjusted to have the same density of particles as
in the canonical ensemble.

This prediction is plotted as a dashed line in Fig.\ref{fig:nv}. While
it is able to correctly reproduce the linear behavior of the canonical
result in the high temperature regime, it strongly deviates from it  
at low temperature: the activation law observed in
the simulations is then replaced in the grand canonical ensemble by a
quadratic dependence on $T$. 
As we shall see in what follows, this deviation is due to the presence
of a condensate, and is similar to the one predicted in \cite{JeanLLL}
for a rotating two-dimensional ideal Bose gas in the lowest Landau level.
Of course, this pathology of the grand canonical ensemble can be
eliminated by a canonization procedure for the vortex density, 
as explained in~\cite{JeanLLL}. 
We give here only the resulting formula:
\begin{equation}
\left(n_{v,+}\right)_{\rm C}=
\frac{m}{4\pi \hbar^2}
\frac{\int_0^{2\pi} d\theta\, e^{-i\theta N} B(\theta) \frac{\sum_\kk
    E_k {\tilde n}_k(\theta)}{\sum_\kk {\tilde n}_k(\theta)}}
{\int_0^{2\pi} d\theta\, e^{-i\theta N} B(\theta)},
\label{eq:nv+can}
\end{equation}
where the generating function $B(\theta)$ is written as
\begin{equation}
B(\theta) = \prod_\kk {\tilde n}_k(\theta)
\end{equation}
in terms of a modified Bose law
\begin{equation}
{\tilde n}_k(\theta) = \frac{1}{e^{\beta(E_k-\mu)}+e^{i\theta}}.
\end{equation}
As one can see in Fig.\ref{fig:nv}, the predictions of
this formula, are in perfect agreement with the
simulation results for 
$g_0=0$.

A physical understanding of the strong suppression of vortices in
the ideal gas when a condensate is present can be obtained by means of
the following approximate treatment based on the Bogoliubov assumption
that the fluctuations of the field in the condensate mode are
negligible.
The 2D classical field $\psi$ can then be expanded as:
\begin{equation}
\psi(\rr) = \psi_0 + \sum_{\kk\neq\mathbf{0}} a_\kk \frac{e^{i\kk\cdot\rr}}{L},
\end{equation}
where the condensate amplitude is fixed to the constant value
\begin{equation}
\psi_0=\left(\frac{\langle N_0\rangle_{\rm Bog}}{L^2}\right)^{1/2}=
\left(\frac{N-\langle\delta N\rangle_{\rm Bog}}{L^2}\right)^{1/2}.
\label{eq:tapis}
\end{equation}
Here $\langle N_0\rangle_{\rm Bog}$ is the mean number of condensate particles in Bogoliubov theory and
the mean number of non-condensed particles $\langle \delta
N\rangle_{\rm Bog}$ in Bogoliubov theory is given by
\begin{equation}
\langle \delta N\rangle_{\rm Bog} = \sum_{\kk\neq\mathbf{0}}
\frac{1}{e^{\beta E_k}-1}.
\end{equation}
Each of the $a_\kk$'s is a complex random variable with a Gaussian
distribution~\cite{why}:  
\begin{equation}
P_\kk(\alpha) \propto e^{-|\alpha|^2 \left(e^{\beta E_k}-1\right)}.
\end{equation}
Since the non-condensed part of the field obeys Gaussian
statistics, the calculation of the mean vortex density can be
analytically performed,
\begin{equation}
\left(n_{v,+}\right)_{\rm Bog} = \frac{m}{4\pi\hbar^2} 
\frac{\sum_{\kk\neq\mathbf{0}} \frac{E_k}{e^{\beta E_k}-1}}{\langle
\delta N\rangle_{\rm Bog}} 
\, e^{-\langle N_0\rangle_{\rm Bog}/\langle \delta N\rangle_{\rm Bog}}.
\label{eq:nvb}
\end{equation}
The prediction of this formula is plotted in Fig.\ref{fig:nv} as a
dot-dashed line: the agreement with the exact results is good.
It is apparent that the dramatic suppression of the vortices in the presence
of a condensate originates from the last factor in Eq.(\ref{eq:nvb}), 
which is indeed
exponentially small in the number of condensate particles.
One can note that a similar factor is involved in the expression for the
probability to have an empty condensate mode in the canonical ensemble. 
On the other hand, the anomalously large vortex density
in the grand canonical ensemble can be explained by the fact that the
most probable value for the number of particles in the condensate mode
is zero in this ensemble.

Before concluding this section, it is important to remind that
(\ref{eq:nvb}) is an approximate expression.
A first necessary condition for its validity is that a condensate is
present, which implies $N\gg \langle \delta N\rangle_{\rm Bog}$.
For a large box $L\gg \lambda_{\rm th}$ ($\lambda_{\rm th}$ is here
the thermal de Broglie wavelength $\lambda_{\rm th}^2=2\pi\hbar^2/m k_B
T$), this condition corresponds to 
\begin{equation}
 n\lambda_{\rm th}^2 \gg 2\, \log(L/\lambda_{\rm th}).
\eqname{cond1}
\end{equation}
Another necessary condition for the validity of \eq{nvb} is that the 
configurations of the field with vortices are still well
described by the Bogoliubov model originally derived for a vortex free
field. 
More precisely, Eq.(\ref{eq:tapis}) has to hold also in presence
of vortices, e.g.\ one has to require that the mean number of
non-condensed particles {\sl conditioned} to the presence of a vortex,
say in $\rr=\mathbf{0}$, remains very close to $\langle \delta
N\rangle_{\rm Bog}$. 
This conditional non-condensed number is defined as 
\begin{equation}
\langle \delta N\rangle^{\rm cond} = \frac{\langle \delta[\psi(\rr=0)]
 \sum_{\kk\neq\mathbf{0}} |a_\kk|^2 \rangle} 
{\langle \delta[\psi(\rr=0)] \rangle}
\label{eq:dNc}
\end{equation}
where the expectation value is taken over the exact field distribution,
$\delta$ is the two-dimensional Dirac distribution and the $a_\kk$'s are
the Fourier components of the field. 
Calculating (\ref{eq:dNc}) within Bogoliubov approximation leads to the
validity condition
\begin{multline}
\langle \delta N\rangle^{\rm cond}_{\rm Bog} - \langle \delta
 N\rangle_{\rm Bog} = \left(2\frac{\langle N_0\rangle_{\rm Bog}}{\langle
 \delta N\rangle_{\rm Bog}}-1\right)\times \\
\times \frac{\displaystyle\sum_{\kk \neq \mathbf{0}}
 \left(\frac{1}{e^{\beta E_k} - 1}\right)^2}
{\langle \delta N\rangle_{\rm Bog}} \ll \langle \delta N\rangle_{\rm Bog}.
\eqname{Bogocond}
\end{multline}
In the large box limit $L\gg \lambda_{\rm th}$, this condition reduces
to the simple condition
\begin{equation}
n\lambda_{\rm th}^2\ll \frac{4\pi^2}{A}\, [\log(L/\lambda_{\rm
th})]^3,
\eqname{cond2}
\end{equation} 
where the numerical coefficient $A=\sum_{\mathbf{q}\in{\mathbb
Z}^{2*}}q^{-4}\simeq 6.0268$.
Note that the two conditions \eq{cond1} and \eq{cond2} are well
compatible in the large box limit $L\gg \lambda_{\rm th}$, and define a
finite validity interval for the Bogoliubov formula \eq{nvb}.

\subsection{General analytical model for the vortex density} 
\label{sec:analyt}
In this subsection we provide a physical explanation to the numerical
observation that the vortex density follows an approximate activation
law at low temperature. 
This is done by developing a simple and physically transparent model
whose predictions turn out to be in good quantitative agreement with
the semi-classical simulations presented in section \ref{sec:numerical},
for both the ideal and the interacting cases.

The idea is to look for an approximate field distribution of
the form
\begin{equation}
P_{\rm simple}[\psi] = e^{-\beta U[\psi]}\, \delta(N-||\psi||^2),
\label{eq:fd}
\end{equation}
where $||\psi||^2=dV \sum_{\mathbf{r}} |\psi(\mathbf{r})|^2$,
with a suitably chosen energy functional $U[\psi]$.
As a temperature independent energy functional (e.g.\ the
Gross-Pitaevskii one \eq{weight}) would introduce an unacceptable
cut-off dependence \cite{gap_zero}, we are forced to allow for a temperature
dependence of $U$.

In the ideal gas case, we can reproduce the reasoning of
Sec.\ref{subsec:itce} starting from a different representation of the
infinite temperature density operator,
\begin{equation}
\rho(\tau=0) = \int \mathcal{D}\psi \, \frac{e^{-||\psi||^2}}{N!}
 |N:\psi\rangle 
\langle N:\psi|,
\end{equation}
which comes from the projection of the standard overcompleteness
relation for the Glauber coherent states onto the $N$-particle
subspace. 
Note that $\psi$ now runs over the whole functional space and is no
longer restricted to the unit sphere.
The evolution \eq{sogpe} is then applied to each initial Fock
state; in the $g_0=0$ case, this can be solved analytically.
Taking the field $\psi$ at `time' $\beta$ rather than at time $0$
as integration variable, we can write
\begin{equation}
\rho(\beta) = \int \mathcal{D}\psi \,  P_0[\psi] |N:\psi/||\psi||\rangle
 \langle N:\psi/||\psi||\, |, 
\end{equation}
with the field distribution $P_0[\psi]$ equal to
\begin{equation}
P_0[\psi] = e^{-||\psi||^2} \frac{||\psi||^{2N}}{N!} e^{-\sum_\kk
 |a_\kk|^2 (e^{\beta E_k}-1)}. 
\end{equation}
$a_\kk$ is here the Fourier component of the field $\psi$ on the
normalized plane wave $e^{i\kk\cdot\rr}/V^{1/2}$. 
The $||\psi||$ dependent prefactor allows for fluctuations of
$||\psi||^2$ at most of order $O(N^{1/2})$  around $N$, which, in the
large $N$ limit, is a relatively small quantity as compared to $N$.
By approximating the prefactor with a Dirac delta 
imposing $||\psi||^2=N$ \cite{ocgbt}, we finally obtain the
desired form (\ref{eq:fd}), with the energy functional
\begin{equation}
U_0[\psi] = \sum_\kk |a_\kk|^2 k_B T (e^{\beta E_k}-1).
\label{eq:U0}
\end{equation}
For the eigenmodes of energy $E_k \ll k_B T$, this energy functional
essentially reduces to the non-interacting Gross-Pitaevskii energy
functional, while for the eigenmodes of energy $E_k\gg k_B T$ the
large value of $e^{\beta E_k}$ strongly reduces the  
modulus of $a_\kk$, as required by the Bose law for a quantum field.

This construction can then be heuristically extended to the
interacting case.
Restricting ourselves to relatively high temperatures $k_B T \gg g_0 n$,
we can assume that the modes for which the interaction energy plays a
significant role have an energy $\lesssim g_0 n$ and can be treated within
a classical field treatment.
This amounts to adding the usual interaction term of the
Gross-Pitaevskii energy functional \cite{jpf} to the ideal gas
functional (\ref{eq:U0}):
\begin{equation}
U[\psi] = \sum_\kk |a_\kk|^2 k_B T (e^{\beta E_k}-1) + \frac{g_0}{2} \int d^2\rr\, |\psi|^4.
\end{equation}
As the norm of $\psi$ is fixed to $N$ in (\ref{eq:fd}),
the energy functional $U$ can be rewritten in the more convenient form
\begin{equation}
U[\psi]  = \frac{N}{||\psi||^{2}}  \sum_\kk |a_\kk|^2 k_B T (e^{\beta E_k}-1)
+ \frac{g_0 N^2}{2 ||\psi||^4} \int d^2\rr\, |\psi|^4,
\label{eq:Ue}
\end{equation}
which is invariant under multiplication of $\psi$ by a global factor,
and allows to formally relax the condition $||\psi||^2=N$.

The fact that the formation of vortices at low temperature is an
activated process results from the fact that the minimal value of
$U[\psi]$ for a field with at least one node is strictly larger than the
absolute minimum of $U[\psi]$ (which corresponds to a nodeless
$\psi$). 
The activation energy $\Delta(T)$ is given by:
\begin{equation}
\Delta(T) \equiv \min_{\psi\,\mathrm{with\,a\,node}}U[\psi] -
\min_{\psi\,\mathrm{nodeless}}U[\psi],
\label{eq:defD}
\end{equation}
and its temperature dependence originates from the temperature
dependence of the energy functional $U$. 
In the regime $k_B T \ll \Delta (T)$, the probability to have the field
with at least one node has the activation form:
\begin{equation}
p_{\rm node} \simeq e^{-\Delta(T)/k_B T} 
\frac{\int_{\psi\,\mathrm{with\,a\,node}}\mathcal{D}\psi\, e^{-\beta(U[\psi]-\Delta)}}
{\int_{\psi\,\mathrm{nodeless}} \mathcal{D}\psi\, e^{-\beta U[\psi]}}
\end{equation}
where the fraction in the right-hand side has an entropic origin and is
expected to be a slowly varying function of $T$.

The general strategy to calculate $\Delta$ is what follows.
Assuming without loss of generality that the node is in
$\rr=\mathbf{0}$, the $\kk=\mathbf{0}$ Fourier component
$a_{\mathbf{0}}$ of the Bose field can be expressed in terms of the
other components: 
\begin{equation}
a_{\mathbf{0}}=-\sum_{\kk\neq \mathbf{0}} a_\kk.
\label{eq:cond0}
\end{equation}
The energy functional $U[\psi]$ is then a function of the $a_{\kk\neq
  \mathbf{0}}$ only and can be minimized without having to impose any
  further constraint.

The calculation of $\Delta(T)$ is straightforward in the ideal gas
case. We have to impose that the first order differential of $U[\psi]$
with respect to the $a_{\kk}$'s vanishes, which leads to the
condition~\cite{noteDelta} 
\begin{equation}
a_\kk = a_\mathbf{0} \frac{\Delta/N}{\Delta/N-\eta_k},
\label{eq:cs}
\end{equation}
where $\eta_k=k_B T (e^{\beta E_k}-1)$.
Inserting this equation into (\ref{eq:cond0}) gives a closed equation
for the activation energy,
\begin{equation}
 1=\sum_{\kk\neq \mathbf{0}} \frac{\Delta/N}{\eta_k-\Delta/N}.
 \label{eq:sn}
\end{equation}
A graphical reasoning shows that there exists a unique solution in the 
interval $0<\Delta/N< \eta_{2\pi/L}$, which is the smallest root of
Eq.(\ref{eq:sn}) and thus gives the value of $\Delta$.
In the large box limit $L\gg \lambda_{\rm th}$, one has the analytic
expansion: 
\begin{equation}
\Delta=\frac{N}{\sum_{\kk\neq\mathbf{0}} \eta_k^{-1}}
\left[ 1-\frac{\sum_{\kk\neq\mathbf{0}} \eta_k^{-2}}
{(\sum_{\kk\neq\mathbf{0}} \eta_k^{-1})^2}+\ldots
\right],
\eqname{Delta_GP}
\end{equation}
whose leading term reduces to
\begin{equation}
\Delta\simeq\frac{\pi\hbar^2 n}{m \log(L/\lambda_{\rm th})}
\eqname{delta_GP2}.
\end{equation}
Remarkably, the condition to be in the activation
regime $\Delta\gg k_B T$ is equivalent to the condition \eq{cond1}
for Bose condensation, $N\gg \langle \delta N\rangle_{\rm Bog}$. 
Note also that the leading term in \eq{Delta_GP} coincides with the
activation part of the Bogoliubov result \eq{nvb}. 
The successive term gives a correction to $\Delta$ which is negligible as
compared to $k_B T$ provided that the validity condition \eq{cond2}
for the Bogoliubov theory is satisfied.

In the interacting case, a numerical minimization of $U[\psi]$ in
the subspace of the fields with a node in $\rr=\mathbf{0}$ is performed
with the conjugate gradient method. 
As an initial guess, a $\psi$ with random complex Fourier coefficients
$a_{\kk\neq\mathbf{0}}$ is used. 
We find that the minimizing field $\psi_0$ has a uniform phase and has
a double node in $\rr=0$.
This means that $\psi_0$ may be taken real and corresponds 
to the superposition of two, oppositely charged vortices located in the
origin.

Note that it is possible to reduce the energy $U$ by continuously
transforming this field configuration into a nodeless configuration with
just a dip in the density at $\rr=\mathbf{0}$.
On the other hand, a continuous transformation of this field
configuration into a configuration with a pair of closely spaced opposite charge
vortices corresponds to an increases of the energy $U$.

\begin{figure}[htb]
\includegraphics[width=8cm,clip]{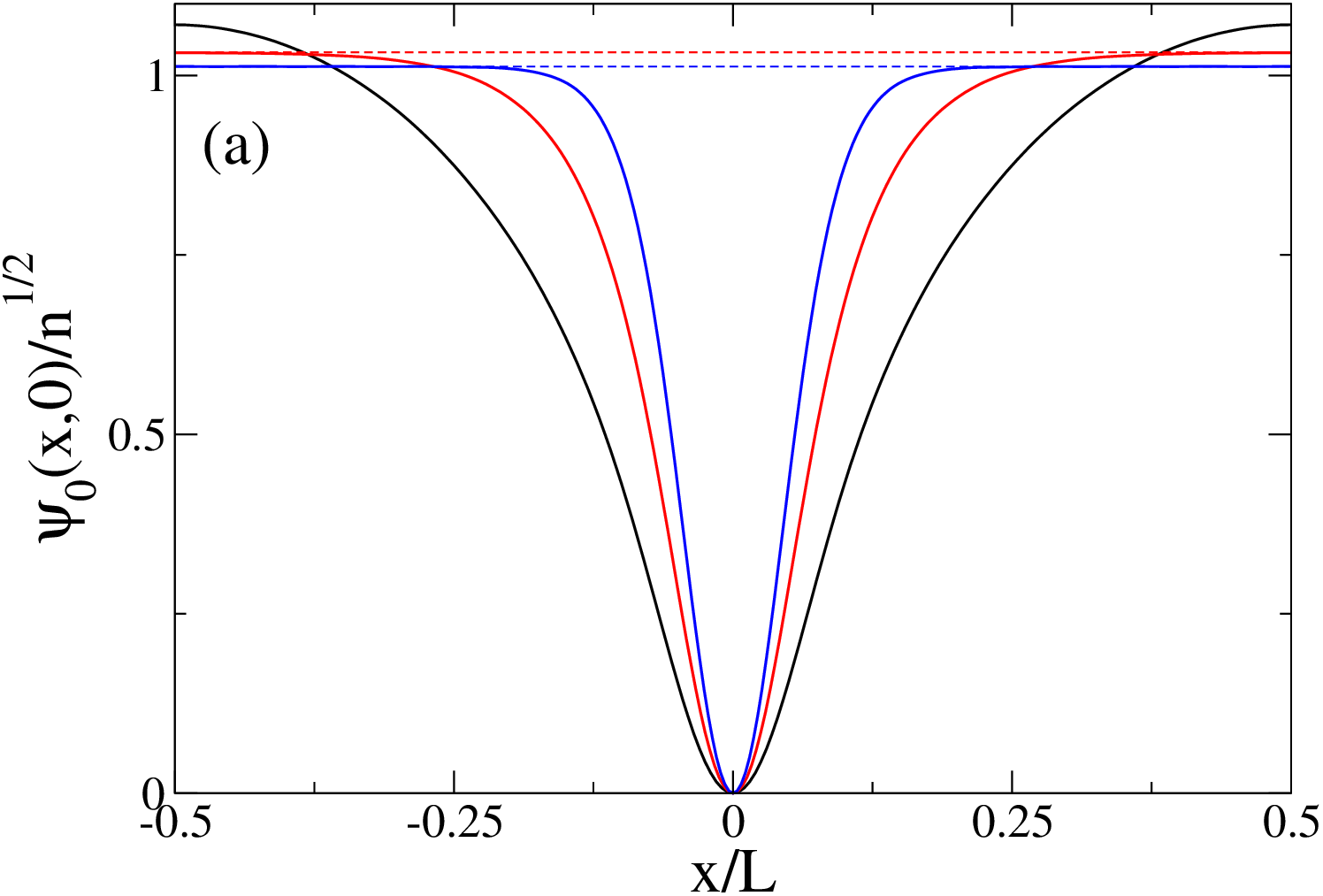}
\includegraphics[width=8cm,clip]{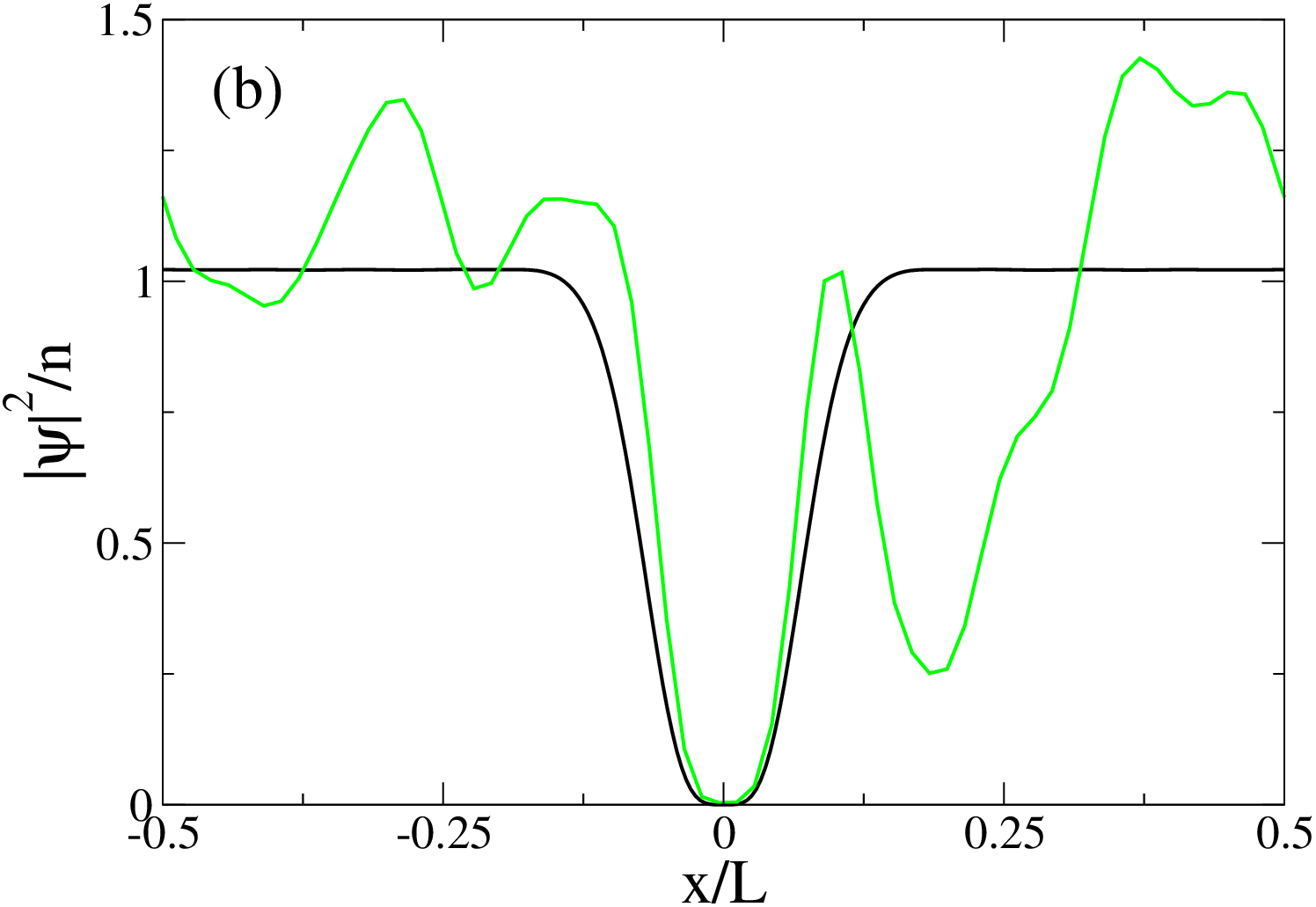}
\caption{(Color online) (a) Cut along $x$-axis of the field $\psi_0$ minimizing the energy
 functional $U[\psi]$ 
over the fields with a node at the origin.
Black solid line (the broadest hole): $g_0=0$, $T/T_d=0.35/(2\pi)\simeq 0.056$;
red solid line: $g_0=0.1 \hbar^2/m$, $T/T_d=0.5/(2\pi)\simeq 0.08$;
blue solid line (the narrowest hole): $g_0=0.333 \hbar^2/m$,
 $T/T_d=0.625/(2\pi)\simeq 0.1$.
The total number of particles is $N=1000$.
The dashed lines for $g_0>0$ correspond to a field value $(\mu/g_0)^{1/2}$, 
where $\mu$ is the Lagrange multiplier defined in Eq.(\ref{eq:nggpe}).
(b) For a semi-classical Monte Carlo realization of the field with a single
vortex pair with a small radius, comparison of the density profile of the field
(green solid line)
with the one of the minimizer $\psi_0$ of $U[\psi]$ with a node (black solid line).
Here $g_0=0.333 \hbar^2/m$, $T/T_d=0.5/(2\pi)\simeq 0.08$,  the vortex
pair diameter is $\simeq 0.03 L$ and the origin of the coordinates was 
redefined to match the location of the vortex pair.
}
\label{fig:trou}
\end{figure}

In Fig.\ref{fig:trou}a we show a cut of the field $\psi_0$ along $x$ axis
for the same parameters as in Fig.\ref{fig:G2v}b. In Fig.\ref{fig:trou}b
we compare the corresponding density profile to the one
of a randomly chosen Monte Carlo realization
with a small radius vortex pair: there
is an acceptable agreement, specially considering the significant density
fluctuations in the simulation result even at the low value of the temperature 
considered here.
It is apparent on
Fig.\ref{fig:trou}a that the field $\psi_0$ has a slowly varying long-distance
tail in the ideal gas case, whereas it rapidly reaches its limiting
value in the interacting case. This can be understood analytically as 
follows.  

For the ideal gas in the thermodynamic limit, one uses 
Eqs.(\ref{eq:cond0}) and (\ref{eq:cs}), neglecting $\Delta/N$ with respect to
$\eta_k$ (for $k\geq 2 \pi/L)$ and then replacing the 
sum over $\kk$ by an integral, to obtain the approximate expression 
\begin{equation}
\psi_0(\rr) \simeq a_{\mathbf{0}} L \frac{\Delta}{N} 
\int \frac{d^2\kk}{(2\pi)^2} \,
 \frac{1-\cos\kk\cdot\rr}{\eta_k},
\end{equation}
which holds  for $r$ much smaller than the box size $L$.
In the limit of large $r\gg \lambda_{\rm th}$,
the integral is dominated by the
contribution of the low momenta, which results in the functional form 
\begin{equation}
\psi_0(\rr) \propto \ln(r/\lambda_{\rm th}).
\eqname{psi0_nint}
\end{equation}

In the interacting case, a sort of generalized Gross-Pitaevskii equation
can be derived, expressing the fact that $\psi_0$ is an extremum of
$U[\psi]$ under the constraint that the norm is constant and a
node is present in $\rr=0$, 
\begin{eqnarray}
\left[k_B T \left(e^{-\beta \hbar^2 \nabla^2/2m}-1\right) +  g_0
 |\psi_0|^2-\mu\right] 
\psi_0(\rr) \nonumber\\ = \left(-\mu L a_\mathbf{0}+g_0\int |\psi_0|^2\psi_0 \right)
\delta(\rr).
\label{eq:nggpe}
\end{eqnarray}
$\mu$ is here the Lagrange multiplier associated to the condition of
a constant norm for $\psi$. 
Using the numerical fact that $\psi_0$ is a real
function and assuming that at large distance from the origin the
laplacian term $\nabla^2 \psi_0$ is negligible, it is easy to see that
$\psi_0^2$ has to converge to the limiting value $\mu/g_0$. 
The normalization condition $||\psi_0||^2=N$ then
leads to $\mu\simeq g_0 n$ in the large $L$ limit.
To see how fast $\psi_0$ reaches its limiting value, we set
$\psi_0(\rr)=(\mu/g_0)^{1/2}[1+\varphi(r)]$ and we linearize the equation
in $\varphi$,
\begin{equation}
\left[k_B T \left(e^{-\beta \hbar^2 \nabla^2/2m}-1\right) + 2\mu\right] \varphi(r)\simeq 0.
\end{equation}
We heuristically assume that, at large $r$, $\varphi$ varies slowly at the scale of the
thermal de Broglie wavelength. 
The first operator in the above equation may then be approximated by the
usual kinetic energy operator, so that 
\begin{equation}
\left[-\frac{\hbar^2 \nabla^2}{2m} + 2\mu\right] \varphi(r)\simeq 0.
\end{equation}
The solution is $\varphi(r)\propto K_0(2r/\xi)$ where $\xi$ is the healing
length, and 
$K_0(u)$ is a Bessel function that tends to zero at large $u$ as
$e^{-u}/u^{1/2}$.  As a consequence, at large $r$,
\begin{equation}
\label{eq:asymp_inter}
\psi_0(\rr) = \left(\frac{\mu}{g_0}\right)^{1/2} \left[1 + O\left(e^{-2r/\xi}\right)\right].
\end{equation}
Since $k_B T \gg g_0 n$, one indeed finds that, at large $r$, $\varphi(r)$ varies slowly
at the scale of $\lambda_{\rm th}$, so that our heuristic assumption is
{\sl a posteriori} justified.

\begin{figure}[htbp]
\includegraphics[width=8cm,clip]{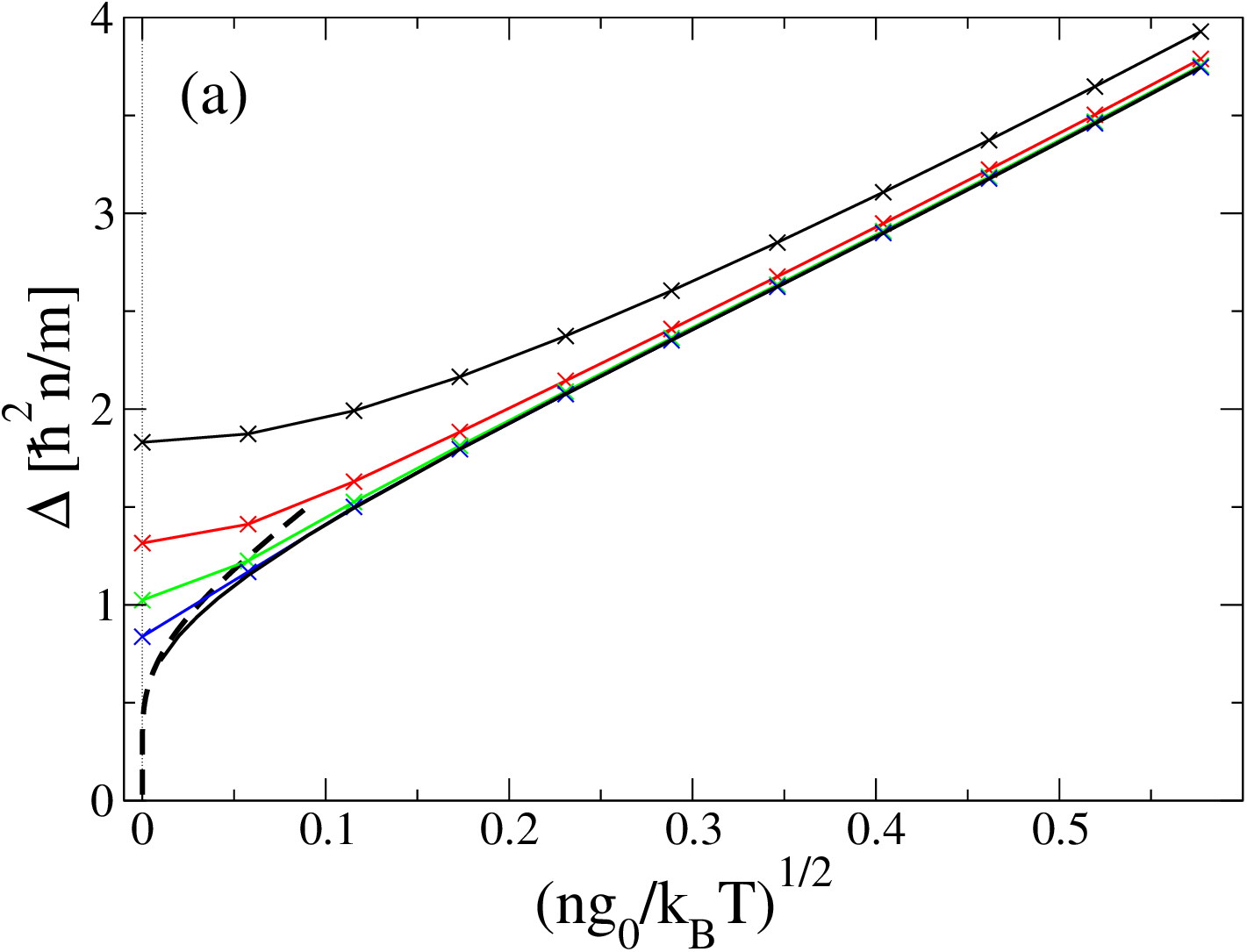}
\vspace{0.5cm}
\includegraphics[width=8cm,clip]{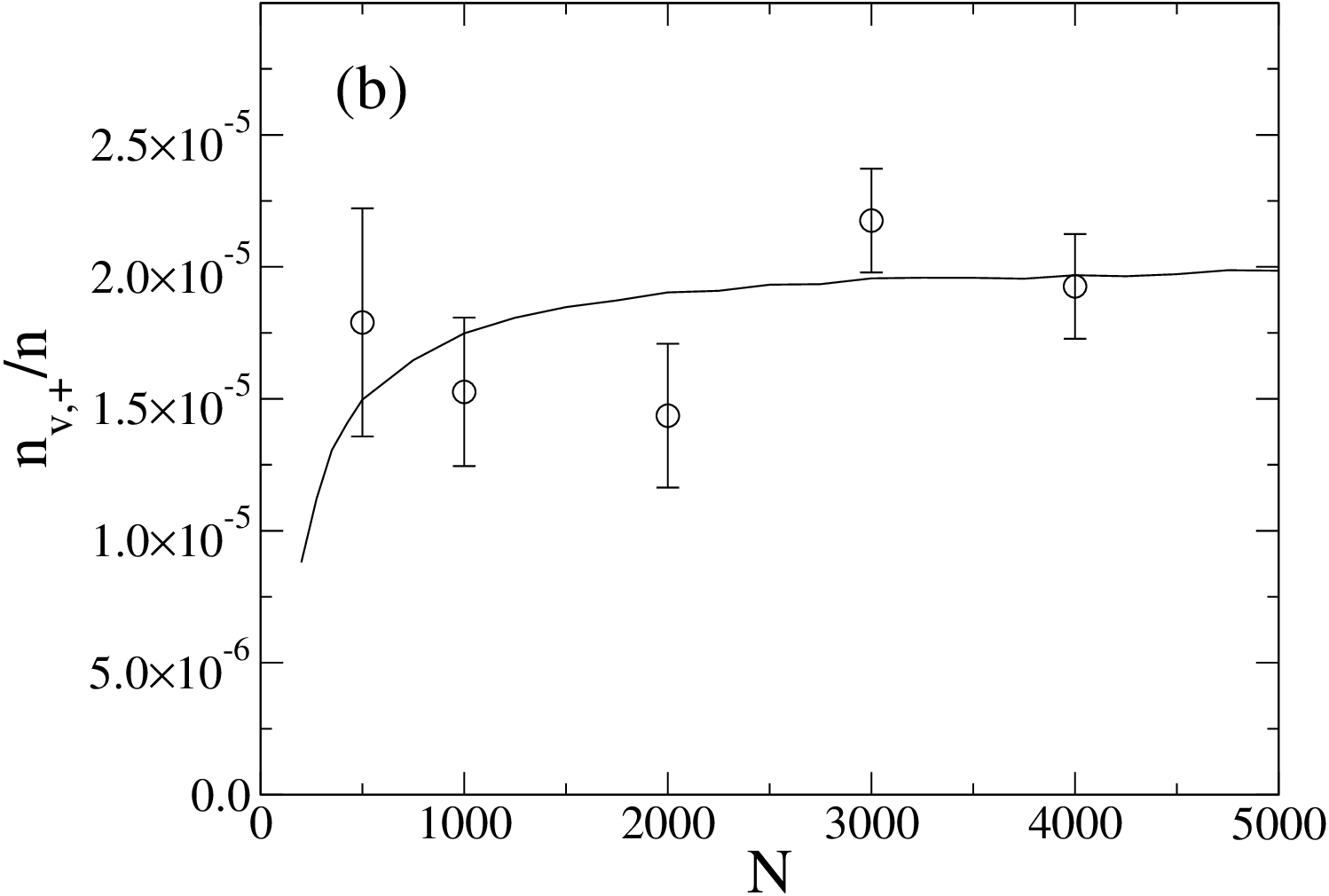}
 \caption{(Color online)
(a) Activation energy $\Delta(T)$ as a function of $(g_0 n/k_B T)^{1/2}$ 
at a fixed particle density $n$ for increasing system size
$L/\lambda_{\rm th}=6,\,12,\,24,\,48$ (thin solid lines, 
respectively black, red,
green, blue, from top to bottom; the crosses are the actually calculated values and the
lines are a guide to the eye).
The dashed line is the upper bound Eq.(\ref{eq:upper}) for an infinite system
size.
The thick solid line is the improved upper bound discussed around
Eq.(\ref{eq:van}), plotted for
$(n g_0/k_B T)^{1/2}\geq 0.01$.
(b) Vortex density as a function of the total particle number
(for increasing system sizes) for fixed values of the density $n$
and the temperature $T=0.5\,T_d/(2\pi)\simeq 0.08\,T_d$, and a coupling constant $g_0=0.333
\,\hbar^2/m$.
Circles: semi-classical simulations. Solid line:
prediction of the activation law
$0.44\, e^{-\Delta/k_B T}$ where the numerical factor $0.44$ 
was fitted to the data.
}
\label{fig:ther_lim}
\end{figure}

This discussion reveals a key difference for the activation energy between
the ideal gas and the interacting gas in the thermodynamic limit. 
While in the ideal gas case the activation energy tends to zero 
in the thermodynamic limit, 
in the interacting case it has a
non-zero limit.
This point is illustrated in Fig.\ref{fig:ther_lim}a, where
we plot the activation energy $\Delta$ as a function of
$(g_0 n/k_B T)^{1/2}$ for increasing system sizes at a fixed particle density
$n$. 
Away from the origin $g_0=0$, a nice convergence towards a universal curve is
obtained, while the dependence of $\Delta$ on the system size remains apparent for
$g_0=0$. 
A physical interpretation of this fact is that, in the interacting case,
the minimizer
$\psi_0$ exponentially converges to a limiting value for
$r\gg \xi$, whereas in the ideal gas case it is
logarithmically sensitive to the box size $L$.

As a consequence of a non-zero value for the activation energy in the thermodynamic
limit, we expect that the vortex density is an intensive quantity for the interacting gas.
This is confirmed by results of Monte Carlo simulations for the vortex density as a 
function of the system size at fixed density and temperature: note on 
Fig.\ref{fig:ther_lim}b how the
vortex density is remarkably constant in the thermodynamic limit.

As is apparent in Fig.\ref{fig:ther_lim}a, the convergence of
the activation energy $\Delta$ to its thermodynamic limit value is not uniform
in $n g_0/k_B T$ but becomes slower and slower for smaller interaction strength.
Analytical results can be obtained for an infinite size system,
as detailed in the appendix \ref{appen:ub}: One finds an upper bound on the
thermodynamic limit value $\Delta_\infty$ of the activation energy,
\begin{equation}
\Delta_\infty \leq \frac{2\pi\hbar^2 n}{m} \, \frac{1-2n g_0/k_B T}{\ln[k_B T/(2 n g_0)]}.
\label{eq:upper}
\end{equation}
This explicit upper bound is represented by a dashed line in Fig.\ref{fig:ther_lim}a.
It shows that $\Delta_\infty$ tends to zero for vanishing interaction strength,
which makes a physical link with the ideal gas result Eq.(\ref{eq:delta_GP2}) in the thermodynamic
limit $L/\lambda_{\rm th}\to \infty$.

A better upper bound, though requiring some numerics,
is obtained by performing a variational calculation, based on the thermodynamic
limit of the ansatz
\begin{equation}
\psi(\mathbf{r})= \mathcal{N} \sum_{\mathbf{k}\neq \mathbf{0}}
\frac{1-\cos(\mathbf{k}\cdot\mathbf{r})}{\exp(E_k/k_B T_{\rm eff})-1+\alpha},
\label{eq:van}
\end{equation}
where $\mathcal{N}$ is a normalisation factor.
The two variational parameters are an `effective' temperature $T_{\rm eff}$
and $\alpha\geq 0$. 
The physical motivation for this ansatz, as well as the way
to implement it in the thermodynamic limit, are given in the appendix \ref{appen:ub}.
The prediction of this ansatz is shown as a thick solid line 
in Fig.\ref{fig:ther_lim}a: it is almost indistinguishable (on the figure)
from the numerical results for the largest system sizes, except in $g_0=0$
where the numerical results suffer from finite size effects.

The success of this ansatz is due to the fact that it
reproduces in a fairly accurate way the spatial shape of $\psi_0$ both at
short and long distances: 
In the limit $n g_0 \ll k_B T$ the energy minimisation leads to
$T_{\rm eff} \simeq T$ and $\alpha \simeq 1.5\, n g_0/k_B T$. 
At distances $r\ll \xi$ one is then allowed to neglect $\alpha$ in
the denominator of \eq{van}. In this way, one recovers the ideal
gas result \eq{psi0_nint} and, in addition, one obtains the 
normalization factor which depends on the interaction strength,
\begin{equation}
\psi_0(\rr) \sim \frac{2\ln(r/\lambda_{\rm th})}{\ln(1/\alpha)}.
\label{eq:asympt_psi0}
\end{equation}
In the large $r$ limit $r\gg \xi$, the ansatz reproduces the
exponentially fast convergence of $\psi_0$ towards its limiting value,
Eq.(\ref{eq:asymp_inter}),
with a decay length differing from the exact one by a numerical
factor close to unity, $\simeq 1.15$.

 From Eq.(\ref{eq:asympt_psi0}), it is possible to estimate the half-width 
 at half maximum of the hole in the density profile $\psi_0^2$:  
in the $g_0 \to0$ limit, a result growing as
$\lambda_{\rm th}\,(\xi/\lambda_{\rm th})^{1/\sqrt{2}}$ is found. 
This prediction is in
good agreement with the numerical results of Fig.\ref{fig:ther_lim}a
for $g_0>0$ and the largest sample size,  $L=48 \lambda_{\rm th}$. 

\section{conclusions}
\label{sec:Conclu}

In this paper, we have introduced a semi-classical field method for the 
study of the thermal equilibrium state of an ideal or weakly interacting
Bose gas at finite temperature.
We have validated the method by verifying that it does not suffer from
ultraviolet divergences and it provides quantitatively accurate predictions
as long as the temperature is higher than the chemical potential of the
gas.
The method being based on a probability distribution in the functional
space of c-number wavefunctions, it appears as being
particularly well suited to the study of thermal vortices,
in contrast to standard Quantum Monte Carlo techniques.

As a first application of the method to a system of current experimental
interest, we have calculated in this paper the density of thermal
vortices in a spatially homogeneous, two-dimensional Bose gas at
thermal equilibrium and we have characterized the spatial correlations
between the positions of 
opposite-charged vortices.
The numerical results are then used as a starting point to develop
simple analytical models and obtain an insight in the physics of the
system in the different regimes.

In both the ideal and the interacting cases, in the low temperature
limit, the vortex density depends on
temperature according to an activation law of the form $\exp(-\Delta/k_B
T)$, with an activation energy $\Delta$ weakly dependent on temperature.
For the ideal gas,  $\Delta$ is non-zero for a finite size system, because
Bose-Einstein condensation takes place in such a system at low
enough temperature; for the same reason, $\Delta$ depends
on the system size and tends logarithmically to zero 
in the thermodynamic limit.
For the interacting gas, $\Delta$ has a non-zero value in the thermodynamic limit,
reached for a system size larger than the healing length $\xi$; this thermodynamic limit
value of $\Delta$ tends to zero logarithmically in the limit of a vanishing interaction
strength.

Finally, we have studied
the spatial correlations between the positions of vortices.
At high temperatures, no qualitative difference appears between the
ideal and the interacting cases.
On the other hand, at low temperatures (i.e.\ in the activation regime),
the correlations have a much longer range
in the ideal gas, which corresponds to the existence of larger size
vortex pairs.

\acknowledgments
We acknowledge the contribution of Bruno Durin and Carlos Lobo in developing a
code for locating the vortices at an early stage 
of this work. We acknowledge useful discussions with Jean Dalibard,
Markus Holzmann, Zoran Hadzibabic, and David Hutchinson.

\appendix

\section{Quantum and semi-classical Glauber-P distributions for a single Bogoliubov mode}
\label{appen:omm}

In this appendix we calculate the exact Glauber-P distribution $P(\gamma)$
and its semi-classical approximation $P_{\rm SC}(\gamma)$
for the thermal density operator of a single mode Hamiltonian of the
form \eq{Hquad}.

The imaginary-time evolution of the Glauber-P distribution
$P(\gamma)$ is very similar to the one \eq{FP} of the full
many-body Hamiltonian:
\begin{multline}
\partial_\tau P(\gamma)=
-E_1(\gamma)\,P(\gamma) \\ 
-\Big\{ \partial_{\gamma}
\left[F_1(\gamma)\,P(\gamma)\right] 
+\frac{\mu}{4}\,\partial_\gamma^2 P(\gamma) +\textrm{c.c.}\Big\}.
\end{multline}
In the $(x,y)$ variables defined as the real and the imaginary parts
of the field $\gamma=x+iy$, 
the mean-field energy $E_1(\gamma)$ and the drift force have the simple
form:
\begin{eqnarray}
E_1(\gamma)&=&(E_k+2\mu)\, x^2+E_k \,y^2 \\
F_1(\gamma)&=&-\frac{1}{2}\left[(E_k+2\mu)\,x + i E_k y \right],
\end{eqnarray}
while the diffusion matrix is non-positive definite due to the 
squeezing terms $\hat{c}^2$ and $(\hat{c}^\dagger)^2$ in the
Hamiltonian.

The analysis of the exact $P(\gamma)$ is most easily done by looking at its Fourier
transform, i.e.\ the normally ordered characteristic function \cite{quantum_noise}:
\begin{equation}
\chi_P(\xi)=\langle e^{\xi\chd}\,e^{-\xi^*\ch} \rangle,
\end{equation}
where the expectation value is taken on the normalized thermal density operator.
For a normalized Gaussian density operator originating from the imaginary-time evolution
under a quadratic Hamiltonian such as \eq{Hquad}, Wick theorem implies
that:
\begin{equation}
\chi_P(\xi)=\exp \left[
\frac{1}{2}\left(\xi^2\,\langle\chd\chd\rangle+
\xi^{*2}\,\langle\ch\ch\rangle-
2\,|\xi|^2\,\langle\chd\ch\rangle\right)\right].
\end{equation}
 From the Gaussian structure of $\chi_P(\xi)$, it is immediate to see
that the Glauber-P distribution is positive and regular if and only if:
\begin{equation}
\langle \chd \ch \rangle > \left| \langle \ch \ch \rangle \right|.
\end{equation}
Applying this to the Bogoliubov theory 
provides the condition on the thermal mode occupation:
\begin{equation}
n_k=\frac{1}{e^{\beta\epsilon_k}-1} > \frac{1}{2}\left[ \left(
  \frac{E_k+2\mu}{E_k}\right)^{1/2}-1\right],
\end{equation}
which is plotted in Fig.\ref{fig:T_min}.

We now turn to the semi-classical approximation.
The solution of the evolution of $\gamma$ under the drift force, Eq.(\ref{eq:phi}),
is a simple scaling transformation: 
\begin{eqnarray}
x(\beta)&=&e^{-(E_k+2\mu)\beta/2}\, x(0) \eqname{xtau} \\
y(\beta)&=&e^{-E_k \beta/2}\, y(0). \eqname{ytau} 
\end{eqnarray}
An explicit form of the weight ${\mathcal W}(x(0),y(0);\beta)$ is
then obtained by inserting the explicit solution
(\ref{eq:xtau}-\ref{eq:ytau}) into \eq{W} and integrating it.
The result is a Gaussian distribution as a function of $(x(0),y(0))$:
\begin{multline}
{\mathcal W}(\beta)=\exp\{-[ (1-e^{-\beta (E_k+2\mu)})\,x(0)^2 \\
+ (1-e^{-\beta E_k})\,y(0)^2]\}.
\eqname{W2}
\end{multline}
The semi-classical approximation \eq{rho_SC_gcan} to the (unnormalized) 
Glauber-P distribution is finally obtained by simply writing \eq{W2}
in terms of the final variables $(x(\beta),y(\beta))$ for which  
$\gamma=x(\beta)+iy(\beta)$.
As the Jacobian of the rescaling transformation (\ref{eq:xtau}-\ref{eq:ytau})
is a constant (independent of $x(0)$ and $y(0)$), 
the result is the Gaussian distribution \eq{gf} with the
widths given by (\ref{eq:srsc},\ref{eq:sisc}).

\section{Numerical algorithm in the canonical ensemble}
\label{appen:nace}

At $\tau=0$, a wavefunction $\psi(0)$ has to be randomly selected on
the unit sphere, and then let evolve until $\tau=\beta=1/k_B T$
according to \eq{sogpe}. This provides the
final value $\psi(\beta)$ of the wavefunction to be used in \eq{rho_SC}.
The observables are then computed as averages over the different
realizations. As \eq{sogpe} is purely deterministic, this reduces to
an averaging over the possible initial wavefunctions $\psi(0)$.

In order to improve the statistical properties of the Monte Carlo
code, an importance sampling technique~\cite{NR} has been implemented
in terms of an {\em 
  a priori} probability distribution $Q[\psi(0)]$.
The expectation value of a generic operator $\hat{O}$ is
rewritten as: 
\begin{equation}
\langle {\hat O} \rangle =\frac{1}{{\mathcal Z}}\,\int_{||\psi||=1}\!{\mathcal
  D}\psi(0)\,Q[\psi(0)]\,
\frac{\langle N:\psi(\beta) | {\hat O} | N:\psi(\beta)\rangle}{Q[\psi(0)]},
\end{equation}
where $\mathcal Z$ is the normalisation factor.
If the distribution of the initial wavefunction $\psi(0)$ is sampled
with a probability law proportional to $Q[\psi(0)]$,
one is left with the average of a quantity
\begin{equation}
\frac{\langle N:\psi(\beta) | {\hat O} | N:\psi(\beta)\rangle
}{Q[\psi(0)]}
\eqname{integrand}
\end{equation}
which can be made flatter by means of a clever choice of
$Q[\psi(0)]$. This provides significant improvement to the statistical
properties of the Monte Carlo code.
In our simulations the form
\begin{equation}
Q[\psi(0)]=\langle N:\psi(\beta)| N:\psi(\beta)\rangle=\|\psi(\beta)\|^{2N}
\end{equation}
is used for the {\em a priori} probability distribution $Q[\psi(0)]$.
This choice was motivated by the requirement that the integrand
\eq{integrand} be flat at least for the calculation of the partition
function, i.e.\ the trace of the density operator.
In the numerical code, the sampling of $Q[\psi(0)]$ is performed by
means of a standard Metropolis algorithm based on rotations in the single-particle
Hilbert space, similarly to what was done in \cite{MCF}.

Another way of sampling $Q[\psi(0)]$ (not used in this work)
could be the following.
At each step of the Metropolis algorithm, a multiplication of the
amplitude of $\psi(0)$ on one (randomly chosen) mode of the system by
a random complex number $z=e^{\lambda} e^{i\alpha}$ is proposed. 
The phase $\alpha$ is uniformly distributed in $[0,2\pi[$ and the
    logarithm $\lambda$ of the modulus has an even probability    
distribution over the real axis.
Subsequently one renormalizes $\psi(0)$. 
One can check that this procedure preserves the detailed
balance condition required by the Metropolis algorithm, since the
probability distributions of $z$ and of $1/z$ coincide.

\section{Upper bound on the activation energy $\Delta$ in the thermodynamic limit}
\label{appen:ub}

In this appendix we derive an upper bound on the thermodynamic limit value
of the activation energy Eq.(\ref{eq:defD}) for an interacting gas $g_0>0$.

To take the thermodynamic limit in the energy functional $U[\psi]$, we set
\begin{equation}
\psi(\mathbf{r}) = \mathcal{N} f(\mathbf{r})
\end{equation}
where $f(0)=0$ and $f(\mathbf{r})$ reaches rapidly unity at large $r/\xi$.
The normalization factor is given by 
\begin{equation}
|\mathcal{N}|^2 \left[L^2 + \int_{L^2} (|f|^2-1)\right] = N,
\label{eq:fnor}
\end{equation}
where we have subtracted and added one to $|f|^2$. As $|f|^2-1$ is an exponentially
narrow function of $r$ for $L\to \infty$, the integral in Eq.(\ref{eq:fnor})
rapidly converges in the thermodynamic limit, so that we get the expansion
\begin{equation}
|\mathcal{N}|^2 = n \left[ 1 - \frac{1}{L^2} \int (|f|^2-1) + O(1/L^4)\right]
\label{eq:nf}
\end{equation}
where the integral is now over the whole plane. This allows to calculate 
the deviation $\delta U_\infty[f]$
between $U[\psi]$ and the nodeless ground state energy $N^2g_0/2L^2$
in the thermodynamic limit, as a functional of $f$. For $U_0[\psi]$ 
the knowledge
of the leading
order term of the normalization factor $\mathcal{N}$
in (\ref{eq:nf}) is sufficient, whereas for the interaction
energy the $1/L^2$ correction is required. We obtain
\begin{eqnarray}
\delta U_\infty[f] &=& n k_B T
\int f^* \left[e^{-\beta\hbar^2\nabla^2/2m}-1\right] f \nonumber \\
&&+\frac{g_0 n^2}{2} \int \left(|f|^2-1\right)^2.
\label{eq:dUinf}
\end{eqnarray}

To easily obtain an upper bound on the thermodynamic limit value $\Delta_\infty$
of the activation energy, we restrict to the class
$\mathcal{C}$ of real and isotropic functions
$f$ such that $0\leq f(\mathbf{r})\leq 1$ for all $\mathbf{r}$. Then $(|f|^2-1)^2= (1-f)^2(1+f)^2
\leq 4(1-f)^2$, so that 
\begin{eqnarray}
\Delta_\infty \leq W[f]&=&n k_B T
\int f^* \left[e^{-\beta\hbar^2\nabla^2/2m}-1\right] f \nonumber \\
&&+2g_0 n^2 \int (1-f)^2,
\ \ \ \forall f\in \mathcal{C}.
\end{eqnarray}
It remains to minimize the energy functional $W[f]$ over the class
$\mathcal{C}$, which is conveniently done in the Fourier
space representation
\begin{equation}
f(\mathbf{r}) = \frac{\int \frac{d^2\mathbf{k}}{(2\pi)^2} u(\mathbf{k})
(1-\cos \mathbf{k}\cdot\mathbf{r})}{\int\frac{d^2\mathbf{k}}{(2\pi)^2} u(\mathbf{k})},
\label{eq:fsr}
\end{equation}
a writing which ensures that $f(0)=0$ and $f\to 1$ at infinity for a smooth
(real) function $u(\mathbf{k})$. This representation leads to
\begin{equation}
W[f] = n\,  \frac{\int\frac{d^2\mathbf{k}}{(2\pi)^2} (\eta_k + 2 n g_0) u(\mathbf{k})^2 }
{\left[\int\frac{d^2\mathbf{k}}{(2\pi)^2} u(\mathbf{k})\right]^2}.
\end{equation}
Imposing that the functional derivative of this expression with respect to $u$ vanishes
leads to the choice 
\begin{equation}
u_m(\mathbf{k}) = \frac{1}{\eta_k + 2 n g_0}.
\label{eq:u_m}
\end{equation}
One can check, at least for $k_B T> 2 n g_0$, that the corresponding function
$f_m(r)$ indeed takes values between 0 and 1 only, so that it belongs to the class
$\mathcal{C}$  and it is the minimizer of $W[f]$ \cite{des}. 
This results in the upper bound \cite{lowb}
\begin{equation}
\Delta_\infty \leq W[f_m] = \frac{2\pi\hbar^2 n}{m} \, \frac{1-2n g_0/k_B T}{\ln[k_B T/(2 n g_0)]}.
\end{equation}

The variational ansatz Eq.(\ref{eq:van}) is deduced from Eq.(\ref{eq:u_m})
by replacing the physical parameters $T$ and $2 n g_0$ by two variational
parameters $T_{\rm eff}$ and $\alpha k_B T_{\rm eff}$.


\begin{thebibliography}{99}
\bibitem{CFT-dyn} Yu. Kagan, B. V. Svistunov, and G. V. Shlyapnikov,
  Sov. Phys. JETP {\bf 75}, 387 (1992);
  Yu. Kagan and B. V. Svistunov, Phys. Rev. Lett. {\bf
  79}, 3331 (1997); N. G. Berloff and B. V. Svistunov, Phys. Rev. A {\bf
  66}, 013603 (2002).

\bibitem{Sachdev}
K. Damle, S. N. Majumdar and S. Sachdev, Phys. Rev. A {\bf 54}, 5037 (1996).

\bibitem{Burnett}
M.J. Davis, S.A. Morgan and K. Burnett, Phys. Rev. Lett. \textbf{87}, 160402 (2001).

\bibitem{Rzazewski0}
K. G\'oral, M. Gajda, K. Rz\c{a}\.zewski, Opt. Express {\bf 8}, 92 (2001);
D. Kadio, M. Gajda and K. Rz\c{a}\.zewski, Phys. Rev. A {\bf 72}, 013607 (2005).

\bibitem{Baym} 	G. Baym, J.-P. Blaizot, M. Holzmann, F. Lalo\"e and
  D. Vautherin, Phys. Rev. Lett. {\bf 83}, 1703 (1999).

\bibitem{3DDeltaT}  P. Arnold and G. Moore,  Phys. Rev. Lett. {\bf 87}, 120401
	(2001); V. A. Kashurnikov, N. V. Prokof'ev, and B. V. Svistunov,
	Phys. Rev. Lett. {\bf 87}, 120402 (2001).

\bibitem{Scalapino} D.J. Scalapino, M. Sears, R.A. Ferrell,
            Phys. Rev. B {\bf 6}, 3409 (1972).

\bibitem{CFT} Y. Castin, R. Dum, E. Mandonnet, A. Minguzzi, I. Carusotto, 
%``Coherence properties of a continuous atom laser'', 
Journal of Modern Optics {\bf 47}, 2671 (2000).

\bibitem{Hutch} T. P. Simula, M. D. Lee, and D. A. W. Hutchinson,
	Phil. Mag. Lett. {\bf 85}, 395 (2005); T. P. Simula and
	P. B. Blakie, Phys. Rev. Lett. {\bf 96}, 020404 (2006).

\bibitem{Svistunov_tc2da}
N. Prokof'ev, O. Ruebenacker, and B. Svistunov, Phys.
Rev. Lett. {\bf 87}, 270402 (2001).

\bibitem{Svistunov_tc2db}
N. Prokof'ev and B. Svistunov,
Phys. Rev. A {\bf 66}, 043608 (2002).

\bibitem{QMC}  D. M. Ceperley, Rev. Mod. Phys. {\bf 67}, 279 (1995).

\bibitem{worm}  M. Boninsegni, N. V. Prokof'ev, and B. V. Svistunov,
	Phys. Rev. E {\bf 74}, 036701 (2006). 

\bibitem{Krauth} W. Krauth, Phys. Rev. Lett. {\bf 77}, 3695 (1996);
% Precision Monte Carlo test of the Hartree-Fock approximation for a trapped Bose gas    
        M. Holzmann, W. Krauth, and M. Naraschewski, Phys. Rev. A {\bf 59}, 2956 (1999);
% Transition Temperature of the Homogeneous, Weakly Interacting Bose Gas    
        M. Holzmann and W. Krauth, Phys. Rev. Lett. {\bf 83}, 2687 (1999).

\bibitem{Ceperley2}
%   Critical Temperature of Bose-Einstein Condensation of Hard-Sphere Gases    
          P. Gr\"uter, D. Ceperley, and F. Lalo\"e, Phys. Rev. Lett. {\bf 79}, 3549 (1997).

\bibitem{stat_N0}
%    Condensate Statistics in One-Dimensional Interacting Bose Gases: Exact Results    
        I. Carusotto and Y. Castin, Phys. Rev. Lett. {\bf 90}, 030401 (2003).

        
\bibitem{Safonov} A. I. Safonov, S. A. Vasilyev, I. S. Yasnikov,
	I. I. Lakashevich, and S. Jaakkola, Phys. Rev. Lett. {\bf 81},
	4545 (1998).

\bibitem{2D_atoms} 
	A. G\"orlitz,  J. M. Vogels, A. E. Leanhardt, C. Raman, T. L. Gustavson, 
        J. R. Abo-Shaeer, A. P. Chikkatur, S. Gupta, S. Inouye, T. Rosenband, and W. Ketterle, 
        Phys. Rev. Lett. {\bf 87}, 130402
	(2001); V. Schweikhard, I. Coddington, P. Engels, V. Mogendorff,
	and E. A. Cornell, Phys. Rev. Lett. {\bf 92}, 040404 (2004);
	D. Rychtarik, B. Engeser, H.-C. N\"agerl, and R. Grimm,
	Phys. Rev. Lett. {\bf 92}, 173003 (2004); N. L. Smith,
	W. H. Heathcote, G. Hechenblaikner, E. Nugent, and C. J. Foot,
	J. Phys. B {\bf 38}, 223 (2005); C. Orzel, A. K. Tuchman,
	M. L. Fenselau, M. Yasuda, and M. Kasevich, Science {\bf 291},
	2386 (2001); S. Burger,  F. S. Cataliotti,  C. Fort,  P. Maddaloni,  
        F. Minardi and  M. Inguscio, Europhys. Lett. {\bf 57}, 1
	(2002); Z. Hadzibabic, S. Stock, B. Battelier, V. Bretin, and
	J. Dalibard, Phys. Rev. Lett. {\bf 93}, 180403 (2004).

\bibitem{Dalib_vort} S. Stock, Z. Hadzibabic, B. Battelier, M. Cheneau,
	and J. Dalibard, Phys. Rev. Lett. {\bf 95}, 190403 (2005).

\bibitem{Dalibard2D} Z. Hadzibabic, P. Kr\"uger, M. Cheneau,
	B. Battelier, and J. Dalibard, Nature {\bf 441}, 1118 (2006).

\bibitem{Cornell_APS} V. Schweikhard, S. Tung, and E. A. Cornell, preprint 
arXiv:0704.0289 [cond-mat.mes-hall].

\bibitem{BKT} V. L. Berezinskii, Zh. Eksp. Teor. Fiz. {\bf 61}, 1144 (1971)
	[Sov. Phys. JETP 34, 610 (1972)];
        J. M. Kosterlitz and D. J. Thouless, J. Phys. C {\bf 5},
	L124 (1972); J. M. Kosterlitz and D. J. Thouless, J. Phys. C {\bf 6},
	1181 (1973); J. M. Kosterlitz, J. Phys. C {\bf 7}, 1047 (1974).

\bibitem{Minnhagen} P. Minnhagen, Rev. Mod. Phys. {\bf 59}, 1001 (1987).

\bibitem{Markus} M. Holzmann, G. Baym, J.-P. Blaizot,
	F. Lalo\"e, PNAS {\bf 104}, 1476 (2007).

\bibitem{XY} M. Le Bellac, {\em Quantum and statistical field theory}
	(Clarendon Press, Oxford, 1991).

\bibitem{Trombetta_2D} A. Smerzi, P. Sodano, and A. Trombettoni,
	J. Phys. B {\bf 37}, S265 (2004).

\bibitem{Trombetta_2D_BEC}
	A. Trombettoni, A. Smerzi, P. Sodano, New J. Phys. {\bf 7}, 57 (2005).

\bibitem{Svistunov2D} Yu. Kagan, V. A. Kashurnikov, A. V. Krasavin,
	N. V. Prokof'ev, and B. V. Svistunov, Phys. Rev. A {\bf 61},
	043608 (2000).

\bibitem{Halperin} B. I. Halperin, Les Houches lecture series,
	eds. R. Balian, M. Kl\'eman and J.-P. Poirier, 
	Vol.35, p. 813 (North-Holland, Amsterdam, 1981).

\bibitem{Berry} M. V. Berry and M. R. Dennis, Proc. R. Soc. Lond. A {\bf
	456}, 2059 (2000).

\bibitem{Glauber} R. J. Glauber, Phys. Rev. \textbf{131}, 2766 (1963); 
R. J. Glauber, Phys. Rev.  Lett. \textbf{10}, 84 (1963).

\bibitem{quantum_optics} D. F. Walls and G. J. Milburn, {\em Quantum Optics}
(Springer-Verlag, Berlin, 1994).

\bibitem{quantum_noise} C. Gardiner and P. Zoller, {\em Quantum Noise}
	(Springer, Heidelberg, 2004).

\bibitem{sim_math} Here $\sim$ is taken in the strict mathematical sense, 
        $f\sim g$ if $\lim f/g=1$.

\bibitem{Leggett} A. J. Leggett, Rev. Mod. Phys. {\bf 71}, S318 (1999).

\bibitem{Svistunov} N. V. Prokof'ev and B. V. Svistunov, Phys. Rev. B {\bf 61},
	11282 (2000).

\bibitem{attention}
Note that the value of $f_n$ predicted by the present Bogoliubov theory
neglects the possibility that the gas populates quasi-condensate states
with a non-zero momentum, that is with non-zero winding numbers along
$x$ or $y$.  
In the thermodynamic limit, this is not a problem in three dimensions.
On the other hand, inclusion of non-zero winding numbers leads to
significant corrections in two dimensions \cite{Svistunov}, 
and dramatically changes the value of $f_n$ in one dimension
\cite{Svistunov,CRAS}.

\bibitem{CRAS} I. Carusotto and Y. Castin, Comptes Rendus Physique
{\bf 5}, 107 (2004).

\bibitem{IYJ_QMC} I. Carusotto, Y. Castin, J. Dalibard, 
%``The $N$ boson time dependent problem: an exact approach with
	%stochastic wave functions'',
Phys. Rev. A {\bf 63}, 23606 (2001).

\bibitem{ShlyapHouches} D. S. Petrov, D. M. Gangardt, and
	G. Shlyapnikov, Lecture notes of Les Houches school on {\sl low
	dimensional quantum gases}, J. Phys. IV France, {\bf 116}, 5
	(2004).

\bibitem{note}
This is correct 
provided that the typical energies per particle (the chemical potential
or the thermal energy scale $k_B T$) do not assume exponentially small values 
$\lesssim \hbar \omega_z e^{-\sqrt{2\pi} a_{\rm ho}/a_{\rm 3D}}$.

\bibitem{Dum2D} 
Y. Castin, R. Dum, Eur. Phys. J. D {\bf 7}, 399-412 (1999).

\bibitem{pas_thermo} Note that the numerical results reported here have
	been obtained for a finite-size system, whose physics is in many
	aspects very different from the one in the thermodynamic limit.
	In this latter case, the non-condensed fraction is in fact one
	for all $T>0$, while the normal fraction tends to zero for $T$
	tending to zero and shows a finite jump at the
	Kosterlitz-Thouless transition temperature \cite{BKT}. In the
	finite systems, the non-condensed fraction is reduced and the
	BKT universal jump smeared out. A discussion of the interplay
	between the Kosterlitz-Thouless transition and Bose-Einstein
	condensation in finite-size 2D systems can be found
	in~\cite{Trombetta_2D_BEC,2D_BEC}.

\bibitem{2D_BEC} S. T. Bramwell and P. C. W. Holdsworth,
	Phys. Rev. B {\bf 49}, 8811 (1994).

\bibitem{Wilkens} M. Wilkens, C. Weiss,  Opt. Express {\bf 1}, 272 (1997).

\bibitem{Olshanii} C. Herzog, M. Olshanii, Phys. Rev. A {\bf 55}, 3254 (1997).

\bibitem{Holthaus}  M. Holthaus and E. Kalinowski, Ann. Phys., NY {\bf
276}, 321 (1999) and references therein.

\bibitem{canon} The (unnormalized) density operator $\rho_{N}$ in the
	canonical ensemble is obtained from the grand canonical one
	$\rho_{GC}$ by projecting this latter onto the subspace with $N$
	atoms. 
        The projector on this subspace ${\mathcal P}_{N}$ can be written
	in the form: 
	${\mathcal P}_{N}=\int_0^{2\pi} d\theta\,e^{i\theta({\hat
	N}-N)}/(2\pi)$, 
	where ${\hat N}$ is the particle number operator. 
	The density operator in the canonical ensemble $\rho_{N}$ 
	is then written
	as the $N$th Fourier harmonic of the grand canonical one
	$\rho_{GC}$ with a complex chemical potential 
	${\tilde \mu}=\mu+i\theta/\beta$:  
	$$\rho_{N}=\int_0^{2\pi}\!\frac{d\theta}{2\pi}\,e^{-\beta
	H}\,e^{(\beta \mu+i\theta){\hat N}}e^{-i\theta N}.$$
	All expectation values can be calculated from this form by means
	of a numerical integration over $\theta$. 

\bibitem{expm} In a fully quantum lattice model, $g^{(2)}(0)$ in 2D
        depends on microscopic aspects of the model, i.e. the lattice spacing;
	this clearly appears in the so-called quantum term in the Bogoliubov theory \cite{Mora}.
        But $g^{(2)}(0)$ as predicted by the semi-classical theory cannot
        capture this microscopic aspect and rather
        has to be interpreted as an extrapolation from the macroscopic
        length scales such as the healing length $\xi$ or the thermal de Broglie
	wavelength $\lambda_{\rm th}$ down to zero, 
        as is apparent in Fig.\ref{fig:g2} for the Bogoliubov gas. 


\bibitem{thermo2} Note that, in the thermodynamic limit, the ideal gas
	is not Bose condensed at any $T>0$, so that $g^{(2)}(0)=2$
	(matter wave Hanbury-Brown and Twiss effect~\cite{HB-T}).
	On the other hand, $g^{(2)}(0)-1$ would be strongly
	reduced
	in an interacting 2D gas~\cite{Safonov,Svistunov2D} at low
	temperature even 
	in the absence of a condensate, the so-called quasi-condensate
	phenomenon~\cite{Popov,ShlyapnikovQC,Stoof,Mora,Sengstock,Aspect}.

\bibitem{ea} In practice, each side of a plaquette is divided in many
	sub-intervals of length $dl$. Over each of them, the field is
	assumed to vary	linearly with the position. The coefficients
	of this linear expansion are deduced from the value of $\psi$ and of
	its derivative at the center of the sub-interval.
	This linear expansion is then used to calculate analytically
        the integral of the
	gradient of the phase of $\psi$ over each sub-interval, and the
	contributions of the different sub-intervals are then summed
	up. 
	It is easy to see why this procedure is much more efficient than
	assuming a linear variation of the phase along each
	sub-interval: while the phase shows rapid variations in the
	vicinity of a vortex core, the variation of the field $\psi$ is
	always smooth.



\bibitem{JeanLLL} Y. Castin, Z. Hadzibabic, S. Stock, J. Dalibard,
	S. Stringari, Phys. Rev. Lett. {\bf 96}, 040405 (2006).

\bibitem{why} Formally, one uses for the non-condensed modes $a_\kk$
	($\kk\neq\mathbf{0}$) a grand canonical Glauber-P distribution
	with a chemical potential $\mu=0$.
	This procedure is justified by observing that the condensate
	mode can be treated as a reservoir of chemical potential $E_0=0$.

\bibitem{noteG2V} It is interesting to remind that in the grand
	canonical ensemble, a simple formula 
	relates~\cite{Berry,Halperin} the charge-weighted vortex pair
	distribution function 
\begin{multline*}
G^{(2)}_{v,Q}(\rr)=2[G^{(2)}_{v,++}(\rr)-G^{(2)}_{v,+-}(\rr)] \\=
\left\langle\,
[\rho_{v,+}(\mathbf{0})-\rho_{v,-}(\mathbf{0})] 
[\rho_{v,+}(\rr)-\rho_{v,-}(\rr)]\,
\right\rangle
\end{multline*}
 to the one-body
	correlation function of the atomic gas 
	$g^{(1)}(\rr-\rr')=\langle \Psihd(\rr')\Psih(\rr) \rangle/n$:
in the thermodynamic limit,
$$
G^{(2)}_{v,Q}(r)=\frac{1}{4\pi^2\,r}\,\frac{d}{dr}
\left( 
\frac{[dg^{(1)}/dr(r)]^2}{1-[g^{(1)}(r)]^2}
\right).
$$

\bibitem{gap_zero} Anticipating on further reasonings,
        one can show for the Gross-Pitaevskii energy functional
        in the absence of cut-off that the activation energy $\Delta$ 
        defined in Eq.(\ref{eq:defD}) is always zero in the thermodynamic
        limit, even for the interacting gas. This results e.g.\ from
        the insertion in Eq.(\ref{eq:dUinf}) of
        a variational ansatz $f(r)$ obtained by setting 
	$u(\mathbf{k})=e^{-k^2 \epsilon}/(1+k^2)$
        in Eq.(\ref{eq:fsr}), and by taking the limit $\epsilon\to 0$.

\bibitem{ocgbt} One can test this approximation, in the interacting case,
        using the fact that the reasoning in \cite{jpf} leads to
	the functional form
	$P[\psi]= ||\psi||^{2N} e^{-||\psi||^2 F[\psi/||\psi||]}/N!$,
	and introducing $\int dN' \delta[N'-||\psi||^2]$ inside the
	functional integral over $\psi$, where $N'$ is
	an intermediate variable. Then the same reasoning as the one exposed
	in the text finds an activation
	energy $\Delta_{\rm bey}=\Delta/(1+\beta g_0 n/2)$.
        Since we are in the limit $n g_0 \ll k_B T$ and $\Delta$ varies
	essentially linearly in $(n g_0/k_B T)^{1/2}$, see Fig.\ref{fig:ther_lim}a,
	we consider this correction to be beyond the accuracy of the simple model
	and we disregard it.

\bibitem{jpf} A more systematic derivation of this form can be obtained
	starting from Eq.(\ref{eq:sogpe}). 
	First, one treats the interaction terms
	to first order in time dependent perturbation theory (here for
	imaginary time). 
	Then, the first order correction is simplified by means of a
	classical field approximation.
	Calculation of the weight $\exp[-||\psi(0)||^2]$ as a function
	of $\psi(\beta)$ finally leads to the energy functional $U$. 

\bibitem{noteDelta} This derivation assumes that $\Delta/N \neq \eta_k$ 
        for all $\kk\neq \mathbf{0}$ in the reciprocal lattice.
	Actually, there exist other solutions $\Delta/N=\eta_q$,
	where $\qq$ is $2\pi/L$ times a non-zero two-dimensional vector with
	integer components. However, they correspond to higher values of
	the energy functional and therefore are not relevant to the present
	discussion.

\bibitem{des} Expanding $u_m(\mathbf{k})$ in powers of
        $\exp(-\beta \hbar^2 k^2/2m)$, one obtains the series expansion 
	$1-f_m(r)=B \sum_{n\geq 0} (1-\alpha)^n e^{-R^2/[2(n+1)]}/(n+1)$
	with $\alpha=2 \beta n g_0$, $B=(1-\alpha)/\ln(1/\alpha)$ and 
	$R^2=r^2 m k_B T/\hbar^2$.

\bibitem{lowb} Let us assume that the minimizer $f_0$ of $\delta U_\infty[f]$
        is a real and increasing function of the distance $r$ to the node location, as suggested
        by the numerical calculations for a finite size system, so that the function $f_0$
        belongs to the class $\mathcal{C}$. Then $(1-f_0^2)^2 \geq (1-f_0)^2$ so that
        $\Delta_\infty\geq W'[f_0]$, where the energy functional $W'$ is deduced from
        $W$ by replacing $2n g_0$ by $n g_0/2$. Since $W'[f_0]\geq \mathrm{min}_{f \in \mathcal{C}}
        W'[f]$, one finds a lower bound
        $\Delta_\infty \geq \frac{2\pi\hbar^2 n}{m} \, \frac{1-n g_0/2k_B T}{\ln[2k_B T/(n g_0)]}$.

\bibitem{NR}  W. H. Press, S. A. Teukolsky, W. T. Vetterling, and B. P.
Flannery, {\rm Numerical Recipes} (Cambridge University Press,
Cambridge, 1988).

\bibitem{MCF}
Y. Castin, I. Carusotto, Opt. Commun. {\bf 243}, 81 (2004).

\bibitem{Mora} C. Mora and Y. Castin, Phys. Rev. A {\bf 67}, 053615 (2003).

\bibitem{HB-T} M. Yasuda, F. Shimizu, Phys. Rev. Lett. {\bf 77}, 3090
	(1996); M. Schellekens, R. Hoppeler, A. Perrin, J. Viana Gomes,
	D. Boiron, A. Aspect, C. I. Westbrook, Science {\bf 310}, 648
	(2005).

\bibitem{Popov} V. N. Popov, {\sl Functional Integrals in Quantum Field Theory
	and Statistical Physics} (Reidel, Dordrecht, 1983).

\bibitem{ShlyapnikovQC} Yu. Kagan, B.V. Svistunov, and G.V. Shlyapnikov,
  Sov. Phys. JETP {\bf 66}, 314 (1987).

\bibitem{Stoof} U. Al Khawaja, J. O. Andersen, N. P. Proukakis, and
	H. T. C. Stoof, Phys. Rev. A {\bf 66}, 013615 (2002).

\bibitem{Sengstock} 
S. Dettmer, D. Hellweg, P. Ryytty, J. J. Arlt,
    W. Ertmer, K. Sengstock, D. S. Petrov, and G. V. Shlyapnikov,
    Phys. Rev. Lett. {\bf 87}, 160406 (2001);
D. Hellweg,
    L. Cacciapuoti, M. Kottke, T. Schulte, K. Sengstock, W. Ertmer,
    and J. J. Arlt, Phys Rev. Lett. {\bf 91}, 010406 (2003).

\bibitem{Aspect} S. Richard, F. Gerbier, J. H. Thywissen, M. Hugbart,
	P. Bouyer, and A. Aspect, Phys. Rev. Lett. {\bf 91}, 010405
	(2003). 

\end{thebibliography}
\end{document}